\documentclass[onecolumn]{aa}
\usepackage{graphics}
\usepackage{epsfig}
\usepackage{txfonts}

\newcommand{\lab}{\label}
\def\beqa{\begin{eqnarray}}
\def\eeqa{\end{eqnarray}}
\def\beq{\begin{equation}}
\def\eeq{\end{equation}}

\def\half{\frac{1}{2}}

\def\gd{g_{\mu\nu}}

\def\umunu{^{\mu\nu}}
\def\dmunu{_{\mu\nu}}

\def\da{_{\alpha}}

\def\uab{^{\alpha\beta}}

\def\ddemunu{_{;\mu\nu}}

\def\ddemu{_{;\mu}}  
\def\ddenu{_{;\nu}}  
\def\ddea{_{;\alpha}}  
\def\ddeb{_{;\beta}}

\let\gam=\gamma
\let\alp=\alpha
\let\sig=\sigma
\let\lb=\label
\renewcommand{\epsilon}{\varepsilon}
\let\no=\nonumber

\def\f{F(\phi)}

\def\p{\phi}

\def\v{V(\phi)}

\begin{document}
\title{Cosmological models in scalar tensor theories of gravity and observations:
a class of general solutions}
\author{M.\,Demianski \inst{1} \fnmsep \inst{2}, E.\,Piedipalumbo\inst{3}\fnmsep \inst{4},
C.\,Rubano\inst{3}\fnmsep \inst{4} , P. Scudellaro\inst{3}\fnmsep
\inst{4} } \offprints{E.Piedipalumbo, ester@na.infn.it}
\institute{Institute for Theoretical Physics, University of
Warsaw,  Hoza 69, 00-681 Warsaw, Poland \and Department of
Astronomy,
 Williams College, Williamstown, MA 01267, USA \and Dipartimento di
Scienze Fisiche, Universit\`{a} di Napoli Federico II, Compl.
Univ. Monte S. Angelo, 80126 Naples, Italy \and Istituto Nazionale
di Fisica Nucleare, Sez. Napoli, Via Cinthia, Compl. Univ. Monte
S. Angelo, 80126 Naples, Italy }
\titlerunning{Scalar tensor theories of gravity and observations}
\authorrunning{M.\,Demianski \& al.}
\date{Received / Accepted}

\abstract{}{To study cosmological models in scalar tensor
theories of gravity with power law potentials as models of an
accelerating universe.}{We consider cosmological models in
scalar tensor theories of gravity that describe an
accelerating universe, and we study a family of inverse power
law potentials, for which exact solutions of the Einstein
equations are known. We also compare theoretical predictions of
our models with observations. For this we use the following
data: the publicly available catalogs of type Ia supernovae and
high redshift Gamma Ray Bursts, the parameters of large scale
structure determined by the 2-degree Field Galaxy Redshift
Survey (2dFGRS), and measurements of cosmological distances
based on the Sunyaev-Zel'dovich effect, among others.}{ We
present a class of cosmological models which describe evolution
of homogeneous and isotropic universe filled in with dust like
matter and a scalar field that is non minimally coupled to
gravity. We show that this class of models depends on three
parameters: $V_0$ - the amplitude of the scalar field potential,
$\widehat{H}_0$ - the present value of the Hubble constant, and
a real parameter $s$ which determines the overall evolution of
the universe. It turns out that these models have a very
interesting feature of producing in a natural way an epoch of
accelerated expansion. We fix the values of these parameters by
comparing predictions of our model with observational data. It
turns out that our model is compatible with the presently
available observational data.}{}
  \maketitle
\section{Introduction}
Recent observations of the type Ia supernovae and CMB anisotropy
indicate that the total matter-energy density of the universe is now
dominated by some kind of dark energy causing an accelerated
expansion of the Universe (\cite{perl,rie+al98, Riess04,spergel}).
The origin and nature of this dark energy remains
unknown~(\cite{zel67,weinberg2}).\\ Prompted by this discovery
recently a new class of cosmological models has been proposed. In
these models the standard cosmological constant $\Lambda$-term is
replaced by a dynamical, time-dependent component - quintessence or
dark energy - that is added to baryons, cold dark matter (CDM),
photons and neutrinos. The equation of state of the dark energy is
assumed to be of a hydrodynamical type $w_{\phi} \equiv \rho_{\phi}
/p_{\phi}$, where $\rho_{\phi}$ and $p_{\phi}$ are, respectively,
the energy density and pressure, and $-1 \leq w_{\phi}<0$, what
implies a negative contribution to the total pressure of the cosmic
fluid. When $w_{\phi} =-1$, we recover the standard cosmological
constant term. One of the possible physical realizations of
quintessence is a cosmic scalar field (\cite{cal+al98}), which
induces dynamically a repulsive gravitational force, that is
responsible for the observed now accelerated expansion of the universe. \\
The existence of dark energy that now dominates the overall
energy density in the universe is posing several theoretical
problems. Firstly, it is natural to ask, why do we observe the
universe at exactly the time when the dark energy dominates over
the matter (\textit{cosmic coincidence} problem). The second
issue, a \textit{fine tuning problem}, arises from the fact that
if the dark energy is constant, like in the standard
cosmological constant scenario, then at the beginning of the
radiation era its energy density should have been vanishingly
small in comparison with the radiation and matter component.
This poses a problem, since in order to explain the inflationary
behaviour of the early universe and the late time dark energy
dominated regime, the dark energy should evolve and cannot
simply be a \textit{constant}. All these circumstances
stimulated a renewed interest in the generalized gravity
theories, and prompted consideration of a variable $\Lambda $
term in more general classes of theories, such as the
scalar tensor theories of gravity.\\

In our earlier paper (\cite{nmc}) we have analyzed extended
quintessence models, for which exact solutions of the Einstein
equations are known, and discussed how in such models it is possible
to treat the fine tuning problem in an alternative way. We applied
our consideration to a special model, based on one of the most
commonly used quintessence potentials $V(\phi)=\lambda \phi^{4}$,
corresponding to the coupling $F(\phi)=(3/32)\phi^2$ (so called
induced gravity). We showed that this model corresponds to a
special, and physically significant case, which emerged by requiring
the existence of a Noether symmetry in the {\em pointlike}
Lagrangian. In this paper we analyze a new and wider class of
theories derived from the Noether symmetry requirement. One of the
main advantages of such models is the fact that they exhibit power
law couplings and potentials, and admit a {\it tracker behaviour}.
In some sense we complete and generalize the analysis initiated in
(\cite{alma, rug1}), where the attention was focused on the
mechanism of obtaining an effective cosmological constant through
the {\it cosmological no hair theorem}, and the analysis of the
solution was restricted to the asymptotical $t\rightarrow \infty$
regime only. Extending our analysis to the whole time evolution, we
are able not only to clarify the properties of such solutions, but
also to compare predictions of such models with observations. We
concentrate on the following data: the publicly available data on
type Ia supernovae and Gamma Ray Bursts, the parameters of large
scale structure determined by the 2-degree Field Galaxy Redshift
Survey (2dFGRS), and  the measurements of cosmological distance with
the Sunyaev-Zel'dovich effect.
\section{Model description}
\subsection{Specifying the model}\label{findingmodel}
Since the detailed properties of a quintessence model, whose
coupling and potential form is derived by requiring the existence of
a Noether symmetry, are extensively discussed in (\cite{nmc}, from
this time on Paper I), here we only summarize the basic results,
referring readers to our previous paper for details. Let us consider
the general action for a scalar field $\phi$, non minimally coupled
with gravity, but not coupled with matter, in this case we have
\begin{equation}
{\cal A} = \int_T \sqrt{-g} \left( F(\phi) R +
\frac{1}{2}g^{\mu\nu} \phi_{, \mu} \phi_{,\nu} - V(\phi) + {\cal
L}_m \right)d^{4}x \,, \label{e1}
\end{equation}
where $F(\phi),~ V(\phi)$ are two generic functions representing
the coupling of the scalar field with geometry and its potential
energy density respectively, $R$ is the scalar curvature,
${\displaystyle \frac{1}{2}g^{\mu\nu} \phi_{, \mu} \phi_{, \nu}}$
is the kinetic energy of the scalar field $\phi$, and ${\cal L}_m$
describes the standard matter content. In units such that $8 \pi
G_{N}=\hbar= c = 1$, where $G_{N}$ is the Newtonian constant, we
recover the standard gravity when $F$ is equal to $-1/2$, while in
general  the effective gravitational coupling is $G_{eff}=-{1\over
{2F}}$. Here we would like to study the simple case of a
homogeneous and isotropic universe, what implies that the scalar
field $\phi$ depends only on time. It turns out that for the flat
Friedman-Robertson-Walker universe filled with matter satisfying
the equation of state $p=(\gamma -1)\rho$ and the scalar field
$\phi$, the action in Eq. (\ref{e1}) reduces to the {\em
pointlike} Lagrangian
\begin{equation}\label{lagrangian}
{\cal L}= 6 F a \dot{a}^2 + 6 F' \dot{\phi}a^2 \dot{a}+
a^3\left({1\over 2} \dot{\phi}^2-V(\phi)\right) - D a^{-3(\gamma
-1)} \,, \label{e2}
\end{equation}
where $a$ is the scale factor and  prime denotes derivative with
respect to $\phi$, while dot denotes derivative with respect to
time. Moreover, the constant $D> 0$ is defined in such a way that
the matter density $\rho_{m}$ is expressed as $\rho_m= D (a_o/
a)^{3\gamma}$, where $1 \leq \gamma \leq 2$. The effective
pressure and energy density of the $\phi$-field are given by
\begin{equation}
p_{\phi}= \frac {1}{2} \dot{\phi}^2- V(\phi)- 2(\ddot{F}+ 2H
\dot{F}) \,, \label{fi-pressure}
\end{equation}
\begin{equation}
\rho_{\phi}= \frac {1}{2} \dot{\phi}^2+ V(\phi)+ 6 H \dot{F} \,.
\label{fi-density}
\end{equation}
These two expressions, even if not pertaining to a conserved
energy-momentum tensor, do define an effective equation of state
$p_{\phi}=w_\phi \rho_{\phi}$, which drives the late time behavior
of the model.

From now on we restrict ourselves to a dust filled universe with
$\gamma=1 $, and $p_m=0$. Using the point like Lagrangian Eq.
(\ref{e2}) in the action and varying it with respect to $\phi$ we
obtain the Euler-Lagrange equations
\begin{equation}
\ddot{\phi}+ 3 H \dot{\phi}+ 6( \dot{H} + 2H^{2})F'+ V' =0\,,
\label{klein-gordon}
\end{equation}
\begin{equation}
2 \dot{H}+ 3 H^2= \frac{1}{2 F} (p_{\phi}+ p_m) \,, \label{e4}
\end{equation}
together with the first integral
\begin{equation}
H^2= -\frac{1}{6 F} \left( \rho_\phi+ \rho_m\right) \,. \label{e3}
\end{equation}

Let us now introduce the concept of an effective cosmological
constant $\Lambda_{eff}$. Using Eq.(\ref{e3}) it is natural to
define the effective cosmological constant as
$\Lambda_{eff}=-{\rho_{\phi}\over {2F}}$ and the effective
gravitational constant as $G_{eff}= -{1\over {2F}}$. With this
definitions we can rewrite Eq.(\ref{e3}) as
\begin{eqnarray}
  3H^2 &=& G_{eff}\rho_m +\Lambda_{eff}\,.
\end{eqnarray}
Introducing the standard Omega parameters by
\[ \Omega_{m}=-{\rho_{m}\over {6FH^{2}}}, \quad
\Omega_{\Lambda_{eff}}={\Lambda_{eff}\over
{3H^{2}}}=-{\rho_{\phi}\over {6FH^{2}}} \,,\] we get as usual that
\begin{eqnarray}
   \Omega_m+\Omega_{\Lambda_{eff}}=1\,.
\end{eqnarray}
Imposing the Noether symmetry in the {\it quintessence
minisuperspace}, $\{a,\phi; \dot{a},\dot{\phi}\}$, where the
point-like Lagrangian is defined, it is possible to exactly
integrate the Einstein field equations Eqs. (\ref{e3}) and
(\ref{e4}), as well as to find a form for the two unknown functions
$F(\phi)$ and $V(\phi))$ (for details see, \cite{rug2, n1}). The
existence of this symmetry actually leads to the following relation
between the functions $F(\phi)$ and $V(\phi)$:
\begin{equation}
V = V_0 (F(\phi))^{p(s)} \,, \label{e5}
\end{equation}
where $V_0$ is a constant and
\begin{equation}\label{ps}
p(s)= \frac{3 (s+1)}{2s +3}\,,
\end{equation}
with $s$  a real number. Moreover, a possible simple choice of the
coupling is
\begin{equation}
F = \xi (s) (\phi+\phi_{0})^2 \,, \label{e6}
\end{equation}
where $\phi_{0}$ is a constants that does not affect our results
and, therefore, from now on we set it to zero, and
\begin{equation} {\displaystyle \xi(s)= \frac{(2 s +3)^2}{48 (s+1) (s+2)}}\,.\label{xi1}
\end{equation}

Let us note that the form of the coupling given by (\ref{e6}) is
quite relevant from the point of view of fundamental physics. It
describes the so called Induced Gravity. The Induced Gravity model
was initially proposed by Zee in 1979 (\cite{Zee}), as a theory of
gravity  incorporating the concept of spontaneous symmetry breaking.
It was based on the observation that in gauge theories the
dimensional coupling constants arising in a low-energy effective
theory can be expressed in terms of vacuum expectation values of
scalar fields. In such a model the gravitational and cosmological
constants are not introduced by hand, but are generated in the
process of symmetry breaking of a scalar field non minimally coupled
with the Ricci scalar in the Lagrangian describing the system.
Once the Noether symmetry is specified it is possible to
find a corresponding conserved quantity and use it as a new
dynamical variable  (for details see (\cite{rug2,n1})). One can then
solve the corresponding Lagrange equations and finally after
returning to the original variables we obtain the sought after
$a(t)$ and $\phi(t)$. The final results can be written in the form
\begin{figure}
\centering{
        \includegraphics[width=5 cm, height=5 cm]{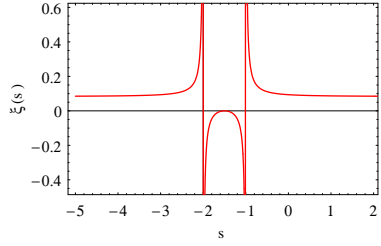}}
        \caption{Diagram of $\xi(s)$ as function of $s$.
        It turns out that an attractive gravity requires $s \in (-2\, , \, \, -1)$.}
        \label{xsiplot}
\end{figure}

\begin{equation}
\label{general2}
 a(t)=A(s)\left(B(s)t^{3\over
 {s+3}}+{D\over {\Sigma_{0}}}\right)^{{s+1}\over
 s}t^{{2s^{2}+6s+3}\over {s(s+3)}}\,,
\end{equation}
\begin{equation}
 \label{generalphi2}
\phi(t)=C(s)\left(-{V_{0}\over \gamma(s)}B(s)t^{3\over
{s+3}}+{D\over \Sigma_{0}}\right)^{-{{2s+3}\over
2s}}t^{-{(2s+3)^{2}\over {2s(s+3)}}}\,,
\end{equation}
where  $A(s)$, $B(s)$, $C(s)$, $\gamma(s)$ and $\chi(s)$ are given
by
\begin{eqnarray}
  A(s)& =&\left( {\chi(s)}\right)^{s+1\over s}  \left({(s+3) \Sigma_0 \over 3\gamma(s) }\right)^{s+2\over
  s+3}\,,\\
  B(s) &=& \left({(s+3)^2 \over s+6}\right)\left({(s+3) \Sigma_0 \over 3\gamma(s) }\right)^{-{3\over (s+3)}}\,,\\
  C(s) &= &\left({\chi(s)}\right)^{-{(2s+3)\over 2 s}}\left({(s+3) \Sigma_0 \over 3\gamma(s) }\right)^{-{(3+2s)\over
  2(s+3)}}\,,
\end{eqnarray}
and
\begin{eqnarray}
  \gamma(s) &=& {2 s+3\over 12 (s+1) (s+2)}\,,\\
  \chi(s)&=& -{  2 s\over 2 s+3}\,,
\end{eqnarray}
where $D$ is the matter density constant, $\Sigma_{0}$ is a constant
of integration resulting from the Noether symmetry, and $V_{0}$ is
the constant that determines the scale of the potential. Together
with the independent parameters, we then use these three constants
$(D, \Sigma_{0}, V_{0}$), which however are not directly measurable,
but they parametrize the possible solutions of the model. In the
next section we shall reduce their number by means of additional
assumptions along the lines of Paper I.

\subsection{The parameter space}
As it is apparent from Eq.(\ref{general2}) and
Eq.(\ref{generalphi2})  there are two additional particular values
of $s$, namely $s=0$ and $s=-3$ which should be treated
independently.

When $s= 0$, a Noether symmetry exists if:
\begin{enumerate}
\item $F= F_0 \phi^2$, and $V = V_0 \phi^2\,,$
\item ${\displaystyle F=-\frac{1}{2}}$ (minimal coupling), and $V
= V_0(A e^{\mu \phi} - Be^{-\mu \phi})^2$, with $\mu
=\sqrt{3/2}$, and $A$, $B$ being constants.
\end{enumerate}
 The case of the minimal coupling has been thoroughly  investigated in
 (\cite{rusc,all03,pv}). If $B=0$ we obtain an exponential potential, which is a very
 important model of quintessence with a standard scalar field.
When $s=-3$ we recover the case of the quartic potential treated in
the Paper I. We shall therefore concentrate on the other values of
$s$.  As we shall see in a moment, this will lead to a very
different class of potentials from those discussed in Paper I. In
fact we obtain inverse power-law type potentials, which are
interesting and recently widely used in the literature.

First of all, we have to find the physically acceptable range for
$s$, and the most important requirement is, of course, that
$G_{eff}>0$, i.e. $F<0$. This restricts $s$ to $(-2, -1)$  as shown
in Fig.(\ref{xsiplot}).

\begin{figure}
\centering{
        \includegraphics[width=5 cm, height= 5 cm]{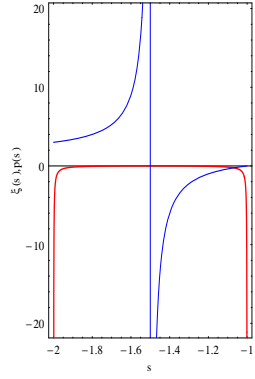}}
        \caption{ Behaviour of the coupling factor $\xi(s)$ (red curve) and the
        power-law exponent $p(s)$ (blue curve): we see that with an appropriate
        choice of s  in the range  $(-1.5\, , \, \, -1)$  all the values for
        the exponents are available.}\label{xpsplot}
\end{figure}

As mentioned above in the range $s \in (-{3\over 2}\, , \,
\, -1)$ the potential $V (\phi)$ is of an inverse power-law type,
$\phi^{-2|p(s)|}$.\,  In this case  our model  naturally admits
cosmological scaling solutions, recently studied, for example by
(\cite{Ame,Uzan}) in the context of quintessence models. In
Fig.(\ref{xpsplot}) we see that all the possible exponents for the
inverse power-law potential are available.\\
To determine the integration constants $D$, and $\Sigma_0$ we follow
the procedure used in Paper I, and we use the age of the universe,
$t_0$, as a unit of time. Because of our choice of time unit, the
expansion rate $H(t)$ is dimensionless, so that our Hubble constant
is clearly of order 1 and is not (numerically) the same as the $H_0$
that is usually measured in ${\rm km s^{-1} Mpc^{-1}}$. Setting $
a_0= a(t_0)=1$ and $\widehat{H}_0=H(t_0)$, we are able to express
$\Sigma_0$ and $D$ as functions of $s$ and $\widehat{H}_0$. We
obtain:
\begin{eqnarray}
  D &=& \left(\left(\frac{1}{A(s)}\right)
   ^{\frac{s}{s+1}}-B(s)\right)
   \Sigma_0 \,,\\
  \Sigma_0 &=& \left(\frac{3^{-\frac{5
   s+6}{s^2+4 s+3}}
   (s+3)^{-\frac{3 s^2+7
   s+3}{s^2+4 s+3}} (s+6)
   \left((\widehat{H}_0-2) s^2+3 (\widehat{H}_0-2)
   s-3\right) \gamma
   (s)^{\frac{s^2-s-3}{s^2+4
   s+3}}}{(s+1) \chi
   (s)}\right)^{\frac{(s+1)
   (s+3)}{s^2-s-3}}.
\end{eqnarray}
Therefore, our family of models,  labeled by $s$, depends only on
$V_0$, and $\widehat{H}_0$. For both these parameters we have not
only a thorough knowledge of their physical meaning, but for
$\widehat{H}_0$ we can also strongly constrain its range of
variability. Actually, we may easily obtain the relation
\begin{equation} \label{ha}
h=9.9{{\widehat{H}_0}\over \tau}\,,
\end{equation}
where as usual  $h=H_{0}/100$ and $\tau$ is the age of the universe
in Gy. We see that $\widehat{H}_0$ fixes only the product $h\tau$.
In particular, we know that $\tau = 13.73^{+ 0.16}_{-0.15}$ (see for instance \cite{spergel}), thus for $\widehat{H}_0
\approx 1$ we get $h< 0.76$. The actual value of $h$ may be
obtained by some of the subsequent tests or in others it has to be
set as a prior - this will be specified in each case considered
below.

\begin{figure}
\centering{
        \includegraphics[width=5 cm, height=5 cm]{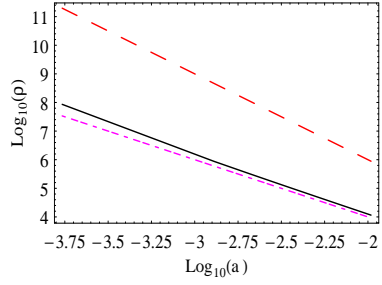}}
        \caption{Plot of $\log_{10}{\rho_{\phi}}$ versus ${\log}_{10} a$ (solid black line).
        The upper and lower dashed lines indicate the log-log plot of $a^{-3}$
        and $a^{-4}$ versus $a$ respectively. It turns out that $\rho_\phi$ scales
        as $a^{-n}$, with $3<n<4$. In this and subsequent plots we use the mean
        values for the parameters obtained through fits (see Table 2).}
        \label{scaling2}
\end{figure}

Using the available observational data we can further constrain the
range of possible values of $s$. Actually requiring that today
$\ddot{a}(t_0)>0$, as indicated by observations of supernovae Ia and
WMAP, we constrain the range of possible values of $s$ to $s \in
(-1.5\, , \, \, -1.2)$.\\
  From the Eqs. (\ref{general2}) and (\ref{generalphi2}) it turns out
that for small values of $t$ the scale factor and the scalar field
behave as
\begin{eqnarray}\label{asymp2}
  a &\propto & t^{{2s^{2}+6s+3}\over {s(s+3)}}, \\
  \phi &\propto & t^{-{(2s+3)^{2}\over
{2s(s+3)}}}.
\end{eqnarray}
Substituting these functions in $\rho_\phi$, as given by
the Eq.(\ref{fi-density}),  we get that for small $t$ the scalar
field density $\rho_\phi \propto a^{-3}$. This is however true
only asymptotically for very small $t$. Exact computation as
shown in Fig. (\ref{scaling2}) gives $\rho_\phi \propto a^{-n}$,
with $n \sim 3$. This justifies our assertion that our model
naturally admits scaling solutions. The situation changes
dramatically near the present time (see below).
\newpage

For large values of $t$ the scale factor and the scalar field
behave as
\begin{eqnarray}
  a &\propto t^{\frac{2 s^2+9 s+6}{s(s+3)}}\,,\\
  \phi  &\propto t^{-\frac{2 s+3}{s}}\,,
\end{eqnarray}

\begin{figure}
\begin{minipage}{13 cm}
\centering{\includegraphics[width=6 cm, height=6
cm]{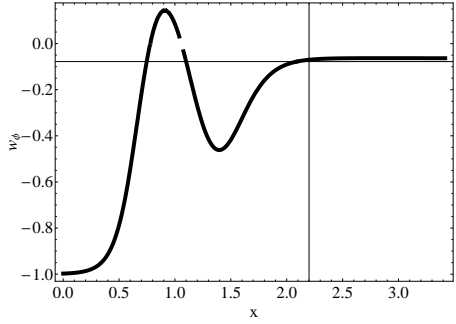}}
        \caption{\small  $w_\phi$ as a function of $x=Log_{10}(1+z)$,
        for the averaged mean values provided by our
        analysis, as shown in Table 2: we observe a  transition from a small constant
        value in the past, $|w_\phi| \approx 0 $,  to $w_\phi = -1$ at present.  }
        \label{wsgen}
\end{minipage}
\end{figure}
\begin{figure}
%\begin{minipage}{10 cm}
\centering{
        \includegraphics[width=5 cm, height=5.5 cm]{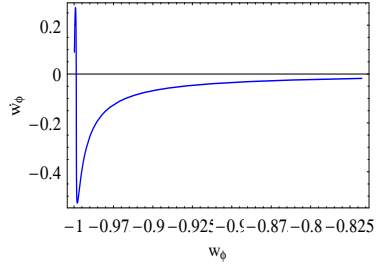}}
        \caption{Rate of change of the equation of state as measured by ${\dot w}_{\phi}$
        versus the $w_{\phi}$ parameter. The values of the parameters
        correspond to the average values  provided by our
        analysis and shown in Table 2.}
        \label{wacc}
%\end{minipage}
\end{figure}

respectively, as shown in (\cite{alma}), where this asymptotical
regime is discussed. It is interesting to note, as shown in
Paper I, that $w_{\phi}$ is representing an equation of state,
in the usual sense, of the effective cosmological constant
$\Lambda_{eff}$, and that $\Lambda_{eff}$ asymptotically behaves
as a true cosmological constant ($w_{\phi}\rightarrow -1$) as $t
\rightarrow \infty$. Let us also mention that since both
$\rho_{\phi}$ and $p_\phi$ depend on $F(\phi)$ through its time
derivative (see Eqs. (\ref{fi-density}) and (\ref{fi-pressure}))
and asymptotically $F(\phi)\sim {\rm constant}$ we  {\it
recover} in this limit the minimally coupled theory. In Figs.
(\ref{wsgen}) and (\ref{wacc}) we show the evolution with the
redshift and the rate of evolution of $w_{\phi}$; the fast
transition toward $w_{\phi}\rightarrow -1$ appears between $z=3$
and $z=0.5$. Before reaching this asymptotic regime the energy
density $\rho_\phi$ is dominated by the coupling term
$6H\dot{F}$. Concluding this section we present the traditional
plot $\log{\rho_{\phi}}$ - $\log a$ and compare it with the
evolution of matter density (see Fig. (\ref{logrhonew})). It is
interesting to see that $\rho_{\phi}$ {\it follows} the matter
density during the matter dominated era, and it becomes dominant
at late time.
\section{Newtonian limit and  Parametrized Post Newtonian (PPN)
behaviour} Recently the cosmological relevance of extended
gravity theories, as scalar tensor or higher order theories, has
been widely explored. However, in the weak field approximation,
all these classes of theories are expected to reproduce the
Einstein general relativity which, in any case, is experimentally
tested only in this limit. This fact is a matter of debate since
several relativistic theories do not reproduce Einstein results at
the Newtonian approximation but, in some sense, generalize them,
giving rise, for example, to Yukawa–-like corrections to the
Newtonian potential which could have interesting physical
consequences. Moreover, in general, any relativistic theory of
gravitation can yield corrections to the Newton potential (see for
example, \cite{will}) which in the post--Newtonian (PPN) formalism
could furnish tests for such theory, mainly based on the Solar
System experiments. In this section we want  to discuss the
Newtonian limit of our class of scalar–-tensor theories of
gravity, the induced gravity theories, and to study the
Parametrized Post Newtonian (PPN) behaviour of these theories.
In particular, it turns out that the Newtonian limit depends on
$\xi(s)$. Furthermore, we find a quadratic correction to the
Newtonian potential strictly depending on the presence of the
scalar--field potential which acts as a cosmological constant.
\subsection{Newtonian limit} In order to
recover the Newtonian limit of our theory described by the action
in Eq. (\ref{e1}), associated with the effective stress-energy
tensor
 \beqa\lb{tensor}
 T^{(eff)}\dmunu= \frac{1}{\f} \left\{ -\half \, \p\ddemu \p\ddenu
 + \frac{1}{4}\, \gd g\uab \p\ddea \p\ddeb\right. &-& \half \, \gd
 \v+ \no\\ -\gd \Box \f &+&\left. \f\ddemunu -{1\over 2}
 \,T^{m}\dmunu \right\},\,
 \eeqa
where $T^{m}\dmunu$ is the usual stress-energy tensor of matter
and $\Box$ is the d'Alembert operator, we write the metric tensor
as
 \beq\lb{metric}
 \gd=\eta\dmunu+h \dmunu\,,
 \eeq
 where $\eta\dmunu$ is the Minkowski metric and
$h\dmunu$ is a small correction. In the same way, we define the
scalar field $\psi$ as a perturbation, of the same order as
$h\dmunu$, of the original field $\p$, that is
 \beq\lb{413}
 \p=\varphi_0+\psi\,,
 \eeq
where $\varphi_{0}$ is a constant of order unity. It is clear that
for $\varphi_{0}=1$ and $\psi=0$ the standard Einstein general
relativity with $\Lambda$ is recovered.

To write the Einstein tensor $G\dmunu$ in an appropriate form, we
define the auxiliary fields
 \beq\lb{metric2}
 \overline{h} \dmunu\equiv h\dmunu-\half\,\eta\dmunu h\,,
 \eeq
and
 \beq\lb{metric3}
 \sig\da\equiv {\overline h}_{\alp\beta,\gam}
 \eta^{\beta\gam}\,,
 \eeq
 where $h=\eta^{\mu\nu} h\dmunu$.
Given these definitions, to the first order in $h\dmunu$, we
obtain
 \beq\lb{einstein}
 G\dmunu=-\half \left\{\Box_{\eta} {\overline
 h}_{\mu\nu}+\eta\dmunu
 \sig_{\alp,\beta}\eta\uab-\sig_{\mu,\nu}-\sig_{\nu,\mu}
 \right\}\,,
 \eeq
where $\Box_{\eta}{\phi}\equiv \eta\umunu \phi_{,\mu \nu}$. To
obtain the weak--field limit of the field equations we have also to
expand the effective stress--energy tensor: this means that it is
necessary to expand the coupling function and the self interacting
potential. Specifically, it turns out that expanding the coupling
function $\f$ and the self--interacting potential $V(\phi)$ (by
using their explicit forms) up to the second order in $\psi$, we get
 \beq\lb{Fexp}
 \f=\xi(s) \left(\varphi_0^2+2\varphi_0\,
 \psi+\,\psi^2\right)\,,
 \eeq
 \beq\lb{Vexp}
 \v\simeq \bar{V}_0 \left(\varphi_0^{p(s)}+p(s)\,\varphi_0^{p(s)-1}\psi+
 \frac{p(s)(p(s)-1)}{2}\varphi_0^{p(s)-2}\psi^2+ \cdots\right)\,,
 \eeq
where  $\bar{V}_0=V_0 \xi(s)^{p(s)}$. Then, to the first order,
the effective stress--energy tensor becomes
 \beq\lb{stressenergy}
 \tilde{T}\dmunu=-2 \varphi_0^{3}\,\eta\dmunu \Box_{\eta} \psi
 +2\, \varphi_0^{3}\psi_{,\mu\nu}-\,\frac{\bar{V}_0\varphi_0^{2+p(s)}}{2\xi(s)}\eta\dmunu
 - {{1\over 2}\frac{\varphi_0^{2}}{\xi(s)}} \,T^{m}\dmunu\,,
 \eeq
and the field equations assume the form
 \beq\lb{fieldexp}
 \half \left\{\Box_{\eta} {\overline
 h}_{\mu\nu}+\eta\dmunu \sig_{\alp,\beta}\eta\uab-\sig_{\mu,\nu}-
 \sig_{\nu,\mu} \right\}=2\,\varphi_0^{3} \eta\dmunu \Box_{\eta} \psi -
 2\,\varphi_0^{3}\psi_{,\mu\nu}+
 \eeq
 $$
 +\,\frac{\bar{V}_0\varphi_0^{2+p(s)}}{2\xi(s)}\eta\dmunu +
 {1\over 2}\frac{\varphi_0^{2}}{\xi(s)}\,T^{m}\dmunu\,.
 $$

 When $T^{m}\dmunu$ describes a point particle of mass $M$ and
 for $p(s)\not= 4$ and $p(s)\not= 1$ we get
\begin{figure}
\begin{minipage}{13 cm}
\centering{
        \includegraphics[width=5 cm, height=5 cm]{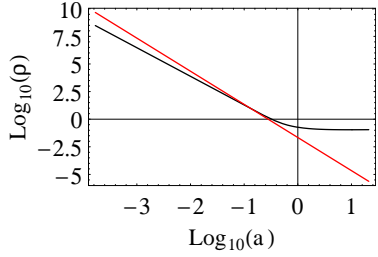}}
        \caption{\small Plot of ${\log}_{10}\rho_{\phi}$ versus ${\log}_{10} a$ in the Jordan frame .
        The vertical bar marks ${\log}_{10}a_0$. The solid red straight
        line indicates the log-log plot of $\rho_m$ versus a. The matter dominated era and the transition to the present dark energy dominated regime are represented. }
        \label{logrhonew}
\end{minipage}
\end{figure}
 \begin{eqnarray}
 h_{00}&\simeq & \left[\varphi_0^{2}\frac{1-16
 \xi(s)}{2\xi(s)(1-12\,\xi(s))}\right]\frac{M}{r}-\left[{M\,\varphi_0^{2}\over 1-12\xi(s)}\frac{\bar{V}_0 (p(s)-4)(p(s)-1)}{1-2\,\xi(s)
 }\right]r\nonumber\\&-& 4\pi
 \left[\frac{ \bar{V}_0\varphi_0^{2+p(s)}}{\xi(s)}+\frac{ 2\varphi_0^{4}}{2 (p(s)-1)}\,\frac{\bar{V}_0 (p(s)-4)(p(s)-1)}{1-2\,\xi(s) }\right]r^2
  \label{428c} \\
  h_{ij} & \simeq & \delta_{ij}\left\{
  \left[\varphi_0^{2}\frac{1-8 \xi(s)}{2\xi(s)(1-12\,\xi(s))}\right]\frac{M}{r}+\left[{M\,\varphi_0^{2}\over 1-12\xi(s)}\frac{\bar{V}_0 (p(s)-4)(p(s)-1)}{1-2\,\xi(s)
 }\right]r \right. \\
&+ & \left. 4\pi \left[\frac{
\bar{V}_0\varphi_0^{2+p(s)}}{\xi(s)}-\frac{ 2\varphi_0^{4}}{2
(p(s)-1)}\,\frac{\bar{V}_0 (p(s)-4)(p(s)-1)}{1-2\,\xi(s)
}\right]r^2\right\}, \label{ 428d}
 \end{eqnarray}
where only terms linear in $V_0$ are given and we omitted the
constant terms.
\subsection{Constraints on PPN parameters}
A satisfactory description of the PPN limit for scalar tensor
theories has been developed in (\cite{esposito-farese, damour2}). In
these papers, this limit has been thoroughly discussed leading to
interesting results even in the case of strong gravitational sources
like pulsars and neutron stars where the deviations from General
Relativity are considered in a non-perturbative regime
(\cite{damour2}). The starting point of such an analysis is a
redefinition of the non minimally coupled Lagrangian action in terms
of a minimally coupled scalar field model {\it via} a conformal
transformation of the form $\tilde{g}_{\mu\nu}\,=
-\,2F(\phi)g_{\mu\nu}$. In fact, assuming the transformation rules:
\begin{equation}\label{conf-phi}
\left(\frac{d\widetilde{\phi}}{d\phi}\right)^2\,=\,\frac{3}{2}\left(\frac{d\ln{F(\phi)}}{d\phi}\right)^2-\frac{1}{2F(\phi)}\,,
\end{equation}
and
\begin{equation}
\widetilde{V}(\phi)\,=\,{V(\phi)\over 4 F(\phi)^{2}}\,, \ \ \ \ \ \
\ \ \ \ \ \ \ \ \widetilde{{\cal L}}_m\,=\,{\cal
L}_m\,F^{-2}(\phi)\,,
\end{equation}
one rewrites the action as
\begin{equation}\label{scatenLag*}
\widetilde{{\cal A}}\,=
\,\int{\sqrt{-\widetilde{g}}\left[\widetilde{R}+\frac{1}{2}\widetilde{g}^{\mu\nu}\widetilde{\phi}_{,\mu}\widetilde{\phi}_{,\nu}-
\widetilde{V}(\widetilde{\phi})+
\widetilde{{\cal L}}_{m}\right]d^{4}x}\,.
\end{equation}
The first consequence of such a transformation is that the
non-minimal coupling is transferred to the ordinary matter sector,
introducing an interaction term between matter and the scalar field.
Actually, the Lagrangian $\widetilde{{\cal L}}_m$ depends not only
on the conformally transformed metric $\widetilde{g}_{\mu\nu}$ and
the matter fields but it also depends on the coupling function.\, In
the same way, the field equations can be recast in the Einstein
frame. The energy-momentum tensor is defined as
$\widetilde{T}^{m}_{\mu\nu}\,=\,\frac{2}{\sqrt{-\widetilde{g}}}\frac{\delta
\widetilde{{\cal L}}_m}{\delta{\widetilde{g}_{\mu\nu}}}$ and is
related to the Jordan expression as $\widetilde{T}^{m}_{\mu\nu}\,=
\sqrt{-2\, F(\phi)}\,T^{m}_{\mu\nu}$. Possible deviations from the
standard General Relativity can be tested through the Solar System
experiments (\cite{will}) and binary pulsar observations which give
an experimental estimate of the PPN parameters. The generalization
of these quantities to the scalar-tensor theories allows the
PPN-parameters to be expressed in terms of the non-minimal coupling
function $F(\phi)$, and
 in our case using the
Eqs. (\ref{e6}) and (\ref{xi1}), we obtain\,:
\begin{equation}\label{gamma}
\gamma^{PPN}-1\,=\,-\frac{F'(\phi)^2}{F(\phi)+2[F'(\phi)]^2}\,=-\,\frac{4\xi(s)}{1+8\xi(s)}\,,
\end{equation}
\begin{equation}
\beta^{PPN}-1\,=\,\frac{F(\phi){\cdot}
F'(\phi)}{F(\phi)+3[F'(\phi)]^2}\frac{d\gamma^{PPN}}{d\phi}\,=0\,.
\end{equation}

\begin{figure}
\centering{
        \includegraphics[width=5 cm, height=7 cm]{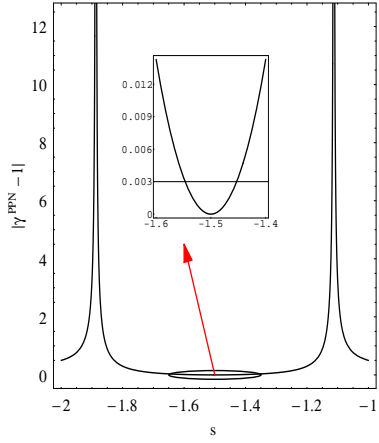}}
        \caption{\small Current limits on the PPN parameters restrict the range of the parameter $s$.
        We see that the constraint on $|\gamma^{PPN}_0-1|< 2\cdot 10^{-3}$ leads to
        $s\in (-1.5,-1.4 )$, as shown in the inner zoom.  }
        \label{gammalim}
\end{figure}

The above definitions imply that the PPN-parameters in general
depend on the non-minimal coupling function $F(\phi)$ and its
derivatives. However in our model $\gamma^{PPN}$ depends only on
$s$ while $\beta^{PPN}=\,1$. The PPN-parameters can be directly
constrained by the observational data. Actually, Solar System
experiments give accurate indications on the ranges of
\footnote{We indicate with the subscript\, $_0$ that the estimates
are based on Solar System observations.} $\gamma^{PPN}_0\,,\
\beta^{PPN}_0$.
\begin{table}[h]
\centering
\caption{\small \label{ppn} A brief summary of recent
constraints on the PPN-parameters.}
\begin{tabular}{|l|c|}
\hline\hline
  Mercury Perih. Shift  (\cite{mercury})&$|2\gamma_{0}^{PPN}-\beta_{0}^{PPN}-1|<3{\times}10^{-3}$ \\\hline
 Lunar Laser Rang.  (\cite{lls})&  $4\beta_{0}^{PPN}-\gamma_{0}^{PPN}-3\,=\,-(0.7\pm 1){\times}{10^{-3}}$ \\\hline
 Very Long Bas. Int.  (\cite{VLBI})&  $|\gamma_{0}^{PPN}-1|\,=\,4{\times}10^{-4}$ \\\hline
 Cassini spacecraft (\cite{cassini})&  $\gamma_{0}^{PPN}-1\,=\,(2.1\pm 2.3){\times}10^{-5}$ \\\hline
\end{tabular}
\end{table}
We summarize the experimental results in Table\ref{ppn}. These
results have been used by (\cite{schimd05}) to set the following
constrains\,:
\begin{equation}\label{gblimits}
|\gamma^{PPN}_0-1|\leq{2{\times}10^{-3}}\,,\ \ \ \ \
|\beta_0^{PPN}-1|\leq{6{\times}{10^{-4}}}.
\end{equation}
It turns out that the limit for $\beta_0^{PPN}$ in the Eq.
(\ref{gblimits}) is naturally verified, for each value of $s$, while
the constraint on $|\gamma^{PPN}_0-1|$ is satisfied only for $s\in
(-1.5,-1.4 )$, as shown in Fig. (\ref{gammalim}).

For the sake of completeness, here we take into account even the
shift that the scalar-tensor gravity induces on the theoretical
predictions for the local value of the gravitational constant as
coming from the Cavendish-like experiments. This quantity
represents the gravitational coupling measured when the Newton
force arises between two masses\,:
\begin{equation}\label{cavendish} G_{Cav}\,=\,\frac{F{\cdot} r^2}{m_1{\cdot} m_2}\,.
\end{equation}
In the case of scalar tensor gravity, the Cavendish coupling is
related to $F$ and $F'$ and is given by\,:
\begin{equation}\label{cavendish-sc}
G_{Cav}\,=\,-\frac{1}{2F(\phi)}\left[1+\frac{[2F'(\phi)]^2}{4F(\phi)+3[2F'(\phi)]^2}\right]\,=
\,-\frac{1}{2\xi(s)\phi^2}\left[\frac{1+16\xi(s)}{1+12\xi(s)}\right]\,
=\,G_{eff}\,\left[\frac{1+16\xi(s)}{1+12\xi(s)}\right]\,,
\end{equation}
and in our models it depends only on $s$

\begin{figure}
\centering{
        \includegraphics[width=5 cm, height=6 cm]{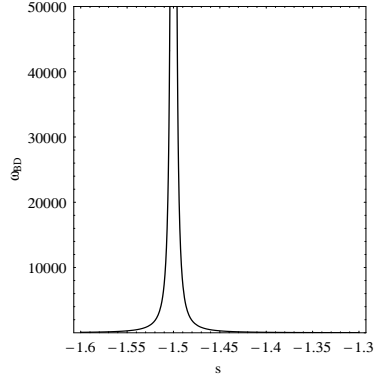}}
        \caption{\small Behaviour of the  Brans-Dicke parameter $\omega_{BD}$
        as a function of $s$. For $s\in (-1.6,-1.3 )$, $\omega_{BD}$
satisfies limits placed by the solar system experiments
($\omega_{BD}>40000$) and by current cosmological observations
($\omega_{BD}>120$)
}.        \label{bransdicke}
\end{figure}

Finally, in Fig. (\ref{bransdicke}) we plot the Brans-Dicke
parameter $\omega_{BD}$ as a function of $s$: actually, for our
model $\omega_{BD}=-\displaystyle\frac{F(\phi)}{2F'(\phi)^2}=
-{1\over 4\xi(s)}$. It turns out that just for $s\in (-1.5,-1.4
)$, $\omega_{BD}$ satisfies the limits coming both from the
Solar System experiments, $\omega_{BD}>40000$ (\cite{will}), and
current cosmological observations, including cosmic microwave
anisotropy data and the galaxy power spectrum data, give
$\omega_{BD}>120$ (\cite{francesca3}).\footnote{ It is worth
noting that the legitimacy of the procedure of direct comparison
of local and cosmological observations, in order to estimate
variations of the physical constants, is a rather strong
assumption, which deserves a proof or at least a
justification. There is actually no reason \textbf{\textit{a priori}}
why local experiments should reveal variations occurring on
cosmological scales, and in regions which are participating in
the Hubble expansion. This interesting aspect of scalar tensor
theories of gravity is discussed in (\cite{barrow1}) and in
(\cite{barrow2}), where it is shown that such procedure is
correct.}

\section{Some considerations about conformal transformations and
interacting dark energy} Conformal transformations  are often used
to convert the non minimally coupled scalar field models into the
minimally coupled ones, to gain mathematical simplification. The
\textit{ Jordan frame}, in which the scalar field is nonminimally
coupled to the Ricci curvature, is mapped into the \textit{ Einstein
frame} in which the transformed scalar field is minimally coupled
but at a price of coupling matter to the scalar field. The two
frames are not physically equivalent, and some care has to be taken
in applying this technique (see for instance (\cite{faraoni2}) for a
critical discussion about this point). In this section we study the
effect of conformal transformations on our models and show that, in
presence of matter, it can mimic a coupling between the quintessence
scalar field and dark matter. We discuss some implications of such a
{\it fictitious} interaction on the effective equation of state.
Actually it turns out that, since the interaction alters the
redshift-dependence of the matter density, it is possible to obtain
an effective transformed dark energy equation of state of
$\widetilde{w}_{\rm eff}<-1$. Let us start from the transformation
rules connected with the conformal transformation
$\widetilde{g}_{\mu\nu}\,= -\,2F(\phi)g_{\mu\nu}$:
\begin{eqnarray}
  {{d\widetilde{\phi}}\over d\phi} &=& \sqrt{{3F'(\phi)^2-F(\phi)\over 2F(\phi)^2}}\label{tranphi}\,, \\
  \widetilde{a} &=& \sqrt{-2 F(\phi)}\,a\label{trana}\,,  \\
  {{d\widetilde{t}}\over dt} &=& \sqrt{-2 F(\phi)}\,,\label{trant}\,\\
 \widetilde{V}(\phi)&=&{V(\phi)\over 4 F(\phi)^{2}}\,,\label{tranv}
\end{eqnarray}
with these new variables the Lagrangian in the
Eq.(\ref{lagrangian}) becomes
\begin{eqnarray}\label{translag}
  \widetilde{{\cal L}} &=&  3\, \widetilde{a}\, \dot{\widetilde {a}}^2 + \widetilde{a}^3\,\left({1\over 2}\,\dot{\widetilde{\phi}}^2-
  \widetilde{V}(\widetilde{\phi})\right) +\widetilde{{\cal L}}_m,
\end{eqnarray}
where now dot denotes derivative with respect to $\widetilde{t}$,
and
\begin{equation}\label{interaction}
\widetilde{{\cal L}}_m (\widetilde{\phi},\widetilde{a})=
\sqrt{-2\, F(\phi)}\, {\cal L}_m = \sqrt{-2\, F(\phi)}\,D
a^{-3(\gamma -1)} .
\end{equation}
An observer in the Einstein frame would infer that the scalar field
is coupled to the dark matter, and this \textit{interaction} is
represented by the term $f(\widetilde{\phi}(\phi))=\sqrt{-2\,
F(\phi)}$, as it can be seen  more clearly from the field equations:
\begin{eqnarray}
  &&\ddot{\widetilde{\phi}}+ 3 \,\widetilde{H}\,\, \dot{\widetilde{\phi}}= - \widetilde{V}_{,\widetilde{\phi}} -{\rho_{m_0}\over \widetilde{a}^3}\,{f_{,\widetilde{\phi}}\over f(\widetilde{\phi}_0)}, \label{traneqs1} \\
&& 3{ \widetilde{H}}^2 =
\left({1\over2}\,\dot{\widetilde{\phi}}^{2}-
  \widetilde{V}(\widetilde{\phi})\right)+ f(\widetilde{\phi})\,{\rho_{m_0}\over f(\widetilde{\phi}_0)}\,\,
  \widetilde{a}^{-3}\label{traneqs2},
\end{eqnarray}
where $\rho_{m_0}$ is the actual matter density (treated here as
dust), $\widetilde{\phi}_0$ the value of $\widetilde{\phi}$ today,
and ${\rho_{m_0}\over f(\widetilde{\phi}_0)}\equiv D$. It turns out
that the density of dark matter does not evolve as
$\widetilde{a}^{-3}$, but scales as $\widetilde{\rho}_m\propto
\displaystyle{f\over \widetilde{a}^{3}}$. Also the Klein-Gordon
equation (\ref{traneqs1}) differs from the standard one because of
the last term on the right hand side.
\begin{figure}
\begin{minipage}{13 cm}
\centering{
        \includegraphics[width=5 cm, height=5 cm]{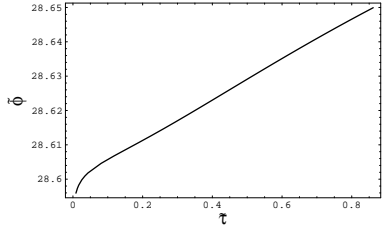}}
        \caption{\small Time evolution of the  transformed scalar field  $\widetilde{\phi}$.  }
        \label{phiconf}
          \end{minipage}
\end{figure}
As shown in (\cite{DAS}) interaction between dark matter and dark
energy could result in an effective equation of state
$\widetilde{p}_{\rm{eff}}=\widetilde{w}_{\rm
eff}{\widetilde{\rho}}_{ \rm{eff}}$, mimicking the $\Lambda$CDM
model. Actually $\widetilde{w}_{\rm eff}$ is defined by the matter
continuity equation
\begin{equation} \frac{d\widetilde{\rho}_{ \rm
eff}}{d\widetilde{t}}=-3\,
\widetilde{H}(1+\widetilde{w}_{eff})\widetilde{\rho}_{ \rm eff}\,,
\label{weffcont}
\end{equation}
where
\begin{eqnarray}\label{effrho}
  \widetilde{\rho}_{\rm{ eff}} \equiv
\frac{\rho_{m_0}}{\widetilde{a}^3}\left(\frac{f(\widetilde{\phi})}{f(\widetilde{\phi}_0)}-1\right)
+ \widetilde{\rho}_\phi\,.
\end{eqnarray}
It can be shown that $\widetilde{w}_{eff}$ is simply related to
$\widetilde{w}_{\phi}$
\begin{eqnarray}
\widetilde{w}_{\rm eff} = {\widetilde{w}_{\phi}\over 1-x}\,,
\label{weff3}
\end{eqnarray}
where
\begin{eqnarray}
x\equiv
-\frac{\rho_{m_0}}{\widetilde{a}^3\widetilde{\rho}_{\phi}}\left(\frac{f(\widetilde{\phi})}{f(\widetilde{\phi}_0)}-1\right).
\label{xdef}
\end{eqnarray}
Since $x=0$ today, one has $\widetilde{w}_{\rm
eff}^{(0)}=\widetilde{w}_{\phi}^{(0)}$, which is greater than or
equal to $-1$. It turns out (\cite{DAS}), however, that   $f$
increases in time, so that $x\geq 0$ and it is possible to have
$\widetilde{w}_{\rm eff}<-1$ in the past. In such a way
interaction between dark matter and dark energy could
generate a superquintessence regime, provided that the
observer treats the dark matter as non-interacting, and ascribes
part of the dark matter density to the scalar field, as is shown
in Eq. (\ref{effrho}). On the other hand an interacting dark
energy component could hide the effect of a non standard
gravity, in the Einstein frame, provided that it is considered
as the physical one.

\begin{figure}
\begin{minipage}{10 cm}
\centering{\includegraphics[width=5 cm, height=5 cm]{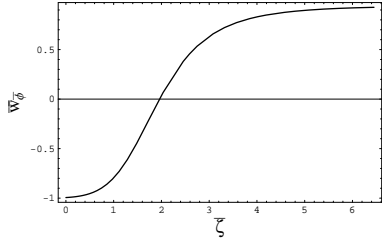}}
        \caption{\small Evolution with the redshift of  $\widetilde{w}_\phi$ in
        the Einstein frame. }
        \label{wtran}
           \end{minipage}
\end{figure}
\subsection{Conformally transformed coupled dark energy }
Let us consider our nonminimally coupled model characterized by the
functions $F(\phi) = \xi (s) \phi^2$ and $V(\phi)= V_0
(F(\phi))^{p(s)}$ respectively. According to the rules in Eqs.
(\ref{tranphi} - \ref{tranv}) we obtain the following relations
between the transformed and original dynamical quantities:
\begin{eqnarray}
  \widetilde{\phi} &=& \sqrt{{12 \xi(s)-1\over 2\xi(s)}}\,\ln{\phi}\,,\label{tranf2} \\
  \widetilde{a} &=& \sqrt{-2\xi(s)}\,\phi\, a\,,\label{trana2}\\
 \widetilde{V} &=&
 {1\over 4}V_0\,\xi(s)^{p(s)-2}\,\phi^{2p(s)-4}\,.\label{tranv2}
\end{eqnarray}
As we see from Eq. (\ref{trana2}), in the transformed frame the
cosmic evolution is \textit{mediated} by the presence of the scalar
field. The explicit form of Eq. (\ref{trant}), which connects the
cosmic time in both frames, can be written in analytical form, but
is indeed rather complicated: actually it turns out that
\begin{eqnarray}\label{trant2}
&&  \widetilde{t} = \left\{2 C(s) s (s+3)
t^{-\frac{2s^2+6s+9}{2(s+3)s}}
\left[\left(\frac{1}{A(s)}\right)^{\frac{s}{s+1}}-B(s)-\frac{V_0
B(s) t^{\frac{3}{s+3}} }{\gamma(s)}\right]^{1-\frac{3}{2 s}}
   \left(1-\frac{V_0 B(s)t^{\frac{3}{s+3}}}{\left(\left(\frac{1}{A(s)}\right)^{\frac{s}{s+1}}-B(s)\right)\gamma (s)}\right)^{\frac{3}{2s}} \gamma
(s)\right\}\nonumber \\
&& \times \,\,{ _2F_1\left[\frac{-2 s^2-6 s-9}{6
s},1+\frac{3}{2s};\frac{-2 s^2-6 s-9}{6s}+1;\frac{V_0
B(s)t^{\frac{3}{s+3}}}{\left[\left(\frac{1}{A(s)}\right)^{\frac{s}{s+1}}-B(s)\right]
\gamma(s)}\right]\over \left(2 s^2+6
s+9\right)\left(\left(\frac{1}{A(s)}\right)^{\frac{s}{s+1}}-B(s)\right)
\left(-\gamma(s)\left(\frac{1}{A(s)}\right)^{\frac{s}{s+1}}+V_0
B(s)t^{\frac{3}{s+3}}+B(s) \gamma(s)\right)}\,.
\end{eqnarray}
From the Eqs. (\ref{tranf2}--\ref{trant2}) we can evaluate the
scalar field energy density and pressure, and the equation of state
$\widetilde{w}_\phi$ according to the \textit{usual} definitions.

As in the Jordan frame, it turns out that also in the Einstein frame
$\widetilde{w}_\phi\rightarrow -1$ for $\widetilde{t}\rightarrow
\infty$ (see for instance Fig. (\ref{wtran})).

\begin{figure}
\begin{minipage}{13 cm}
\centering{
        \includegraphics[width=5 cm, height=5 cm]{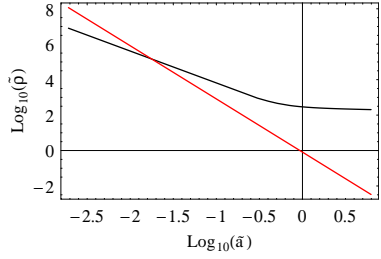}}
        \caption{\small Plot of ${\log}_{10}\widetilde{\rho}_{\phi}$ versus ${\log}_{10}\widetilde{ a}$ in the Einstein frame.
        The vertical bar marks ${\log}_{10}\widetilde{a}_0$. We see that, just
because of the interaction term, $\widetilde{\rho}_{\phi}$ does
not track anymore the matter during the matter dominated era, as
it happens in the Jordan frame (see Fig. \ref{logrhonew}), and
becomes dominant at later times.}
        \label{logrhotra}
        \end{minipage}
\end{figure}

Concluding this section we present the traditional plot
$\log{\widetilde{\rho}_{\phi}}$ - $\log\widetilde{ a}$ and compare
it with $\log{\rho_{m}}$ - $\log{ a}$ relation (see Fig.
(\ref{logrhotra})). Interestingly we see that, just because of the
interaction term, $\widetilde{\rho}_{\phi}$ does not track anymore
the matter during the matter dominated era, as it happens in the
Jordan frame (see Fig. \ref{logrhonew}), but becomes dominant at
earlier times.
 It is interesting to write down the effective equation of
state $\widetilde{w}_{eff}$ (see Eq. \ref{weffcont}), which
mimics a CDM model. Actually in our case $\widetilde{w}_{eff}$
is
\begin{eqnarray}
\widetilde{w}_{\rm eff} =
{\widetilde{w}_{\phi}\over1-x}\equiv-\frac{\rho_{m_0}}{\widetilde{a}^3\widetilde{\rho}_{\phi}}\left(\frac{\widetilde{\phi}+\widetilde{\phi}_0}{2\widetilde{\phi}_0-1}\right)
\,. \label{weff4}
\end{eqnarray}
It turns out that in our case $\widetilde{w}_{eff}$ and
$\widetilde{w}_{\phi}$ are practically indistinguishable,
since in our model the transformed scalar field
$\widetilde{\phi}$ weakly evolves with time as shown in
Fig.(\ref{phiconf})

\subsection{Non minimally coupled quintessence and mass varying
neutrinos through conformal transformations} In this section  we
briefly discuss how the scalar tensor theories of gravity could
be involved in a cosmological model with \textit{mass varying
neutrinos} that mimic the dark energy, a quite different
theoretical scenario of evolution of the universe that recently
has been suggested by (\cite{fardon}). Let us recall that
recently the mass differences between neutrino mass eigenstates
($m_1$, $m_2$, $m_3$) have been measured in oscillation
experiments (\cite{pastor}). Observations of atmospheric
neutrinos suggest a squared mass difference of $\Delta m^2 \sim
3 \times 10^{-3} eV^2$, while solar neutrino observations, and
results from the KamLAND neutrino experiment, point towards
$\Delta m^2 \sim 5 \times 10^{-5} eV^2$. While only weak
constraints on the absolute mass scale ($\Sigma m_{\nu} =
m_1+m_2+m_3$) have been obtained from single $\beta$-decay
experiments, the double beta decay searches from the
Heidelberg-Moscow experiment have reported a signal for a
neutrino mass at $>4\sigma$ level ( \cite{Kl04}), recently
promoted to $>6\sigma$ level  (\cite{Kl06}). This last result
translates into a total neutrino mass of $\Sigma m_{\nu} > 1.2
eV$ at $95 \%$ c.l., but this claim is still considered as
controversial (see \cite{Elli}). It is known in the literature
(\cite{pastor}) that massive neutrinos can be extremely relevant
for cosmology as they leave key signatures in several
cosmological data sets. More specifically, massive neutrinos
suppress the growth of fluctuations on scales below the horizon
scale when they become non relativistic. Current cosmological
data have been able to indirectly constrain the absolute
neutrino mass to $\Sigma m_{\nu} < 0.75$ eV at $95 \%$ c.l.
(\cite{spergel}), and are challenging the Heidelberg-Moscow
claim. However, as first noticed by (\cite{hannestad}), there is
some form of anticorrelation between the equation of state
parameter $w$ and $\Sigma m_{\nu}$. The cosmological bound on
neutrino masses can therefore be relaxed by considering a dark
energy component with a more negative value of $w_{eff}$ than a
cosmological constant. Actually it has been proved that the
Heidelberg-Moscow result is compatible with the cosmological
data only if the equation of state (with $w$ being constant) is
$w_{eff} < -1$ at $95 \%$. This result suggests an interesting
link between neutrinos and dark energy (see for instance
\cite{brook1,brook2,followups,kaplan, amendola2}). According to
this scenario the late time accelerated expansion of the
universe is driven by the coupling between the quintessential
scalar field and neutrinos. Due to such coupling the mass of
neutrinos becomes a function of this scalar field. Since the
scalar field evolves with time the mass of neutrinos is not
constant (mass-varying neutrinos): the main theoretical
motivation for such connection relies on the fact that the
energy scale of the dark energy is of the order of the neutrinos
mass scale.  Moreover, as discussed above, in interacting dark
energy models, the net effect of the interaction is to change
the apparent equation of state of the dark energy, allowing a so
called superquintessence regime, with $w_{eff}<-1$.
Interestingly enough, if the Heidelberg-Moscow results are
combined with the WMAP 3-years data and other
independent cosmological observations, as the ones connected
with the large scale structure -- coming from galaxy redshift
surveys and Lyman-$\alpha$ forests, or with the SNIa surveys,
it is possible to constrain the equation of state to $-1.67<
w_{eff} <-1.05$ at $95 \%$ c.l., ruling out a cosmological
constant at more than $95 \%$ c.l.  (see \cite{melchiorri}).

In the following we will discuss the coupling between neutrinos
and dark energy from the point of view of the conformal
transformation, according to the arguments outlined in the
previous section, i.e. we will show that the neutrino mass and
scalar field coupling can be interpreted as an effect of
conformal transformations from the Jordan to the Einstein frame,
just as it happens for the coupling between the dark energy and
the dark matter. For our purpose the neutrinos can be either
Dirac or Majorana particles, the only necessary ingredient is
that, according to (\cite{fardon}), the neutrino mass is a
function of the scalar field. In the cosmological context,
neutrinos should be treated as a gas (\cite{brook1}) and
described by the collisionless distribution function
$f(x^i,p^i,\tau)$ in the phase space (where $\tau$ is the
conformal time) that satisfies the Boltzmann equation. When
neutrinos are collisionless the distribution function $f$ does
not depend explicitly on time. We can solve then the Boltzmann
equation and  calculate the energy density stored in neutrinos
($f_0$ is the background neutrino distribution function):
\begin{equation}
\widetilde{\rho}_\nu = \frac{1}{\widetilde{a}^4} \int
\widetilde{q}^2 d\widetilde{q} d\Omega \epsilon f_0(\widetilde{q}),
\end{equation}
with $\epsilon^2 = {\widetilde{q}}^2 + {m_\nu(\widetilde{\phi})}^2
{\widetilde{a}}^2$, $\widetilde{a}$ is the scale factor and
$\widetilde{q}^i=\widetilde{ a}\,\,\widetilde{p}^i $ is the
comoving momentum. The pressure is
\begin{equation}
\widetilde{p}_\nu = \frac{1}{3 \widetilde{a}^4} \int
\widetilde{q}^2 d\widetilde{q} d\Omega f_0(\widetilde{q})
\frac{\widetilde{q}^2}{\epsilon}.
\end{equation}
From these equations we derive that
\begin{equation}\label{neutrinoequation}
\dot{\widetilde{\rho}}_\nu + 3\widetilde{H}(\widetilde{\rho}_\nu +
\widetilde{p}_\nu) = \frac{\partial \ln m_\nu}{\partial
\widetilde{\phi}}\dot{\widetilde{\phi}}(\widetilde{\rho}_\nu -
3\widetilde{p}_\nu)
\end{equation}
(note that here the dot denotes the derivative with respect to
$\tau$). The Klein Gordon  equation for the scalar field reads
\begin{equation}\label{scalarequation}
\ddot{\widetilde{\phi} }+ 2 \widetilde{H}\dot{\widetilde{\phi}} +
\widetilde{a}^2\frac{\partial
\widetilde{V}}{\partial\widetilde{\phi}} = -
\widetilde{a}^2\frac{\partial \ln m_\nu}{\partial
\widetilde{\phi}}(\widetilde{\rho}_\nu-3\widetilde{p}_\nu).
\end{equation}
We see that Eq. (\ref{scalarequation}) is formally equivalent to Eq.
(\ref{traneqs1}), but now the term on the right hand side describes
coupling of the scalar field to the neutrino mass. As shown above
the interaction between neutrinos and dark energy could result in an
effective equation of state $\widetilde{w}_{eff}$, defined in Eq.(
\ref{weffcont}), while $\widetilde{w}_{\phi}$ is defined in the
standard way by
$\widetilde{p}_\phi=\widetilde{w}_{\phi}\widetilde{\rho}_\phi$. As
in the previous section the $\widetilde{w}_{eff}$ parameter is
related to $\widetilde{w}_{\phi}$ by
\begin{equation}\label{eq:wapp}
\widetilde{w}_{eff}=\frac{\widetilde{w}_\phi}{1-\widetilde{x}},
\end{equation}
with

\begin{equation}
\widetilde{x}=-\frac{\widetilde{\rho}_{\nu0}}{\widetilde{a}^3\widetilde{\rho}_\phi}\left[\frac{m_\nu(\widetilde{\phi})}{m_\nu(\widetilde{\phi}_0)}-
1\right].
\end{equation}
Also in this case $\widetilde{w}_{eff}$ can be less than $-1$, as
was pointed out  in the context of models with dark matter/dark
energy interaction. This circumstance could lead to an
observational test to establish which of the frames, Jordan or
Einstein, is the physical one, since the coupling between the
quintessential scalar field is provided by the  function $F(\phi)$,
which should drive not only the dark matter evolution (see Eq.
(\ref{traneqs1})), but also the neutrinos mass variation and the
evolution of the gravitational constant $G_{eff}=-{1\over {2F}}$.
\section{Observational data and predictions of our models}

\begin{figure}
\centering{
\includegraphics[width=6 cm, height=6. cm]{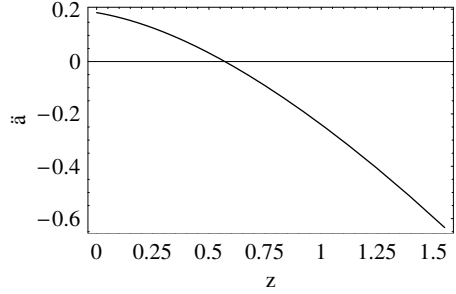}}
 \caption{\small Redshift dependence of the second derivative of the scale
   factor. The transition from a decelerating to an
accelerating expansion occurs close to $z\sim 0.5 $,
 as predicted by recent observations of SNIa $z_t= 0.46\pm 0.13$ (\cite{Riess06,Riess04}).}
\label{transition}
\end{figure}
\subsection{Constraints from recent SNIa observations}

In this section we present results of fits of predictions of our
model to the best SNIa data sets presently available. As a starting
point we consider the sample of 182 SNIa compiled in
(\cite{Riess06}), which includes the 21 new Type Ia supernovae
recently discovered with the {\it Hubble Space Telescope (HST)}, and
combines previous SNIa data sets, namely the Gold Sample compiled in
(\cite{Riess04}) supplemented by the SNLS data set (\cite{SNLS}).
Following the procedure described in Paper I, we perform a
$\chi^2$ analysis comparing the redshift dependence of the
theoretical values to the observational estimates of the distance
modulus, $\mu=m-M$, which in scalar tensor theories of gravity takes
the form
\begin{equation}
m-M=5\log{D_{L}(z)}+  25 + \Delta M_{G}.
\label{eq:modg}
\end{equation}
Here the presence of the correction term
 \begin{equation} \Delta M_{G}={15\over
4}\log\left(G_{eff}\over G_{eff_0}\right),\label{corgef}
\end{equation}
describes the effect of time variation of the effective
gravitational constant $G_{eff}$ on the luminosity of high
redshift supernovae, and allows one to test the scalar tensor
theories of gravity (\cite{Gat01,uz}) using the SNIa data.
Moreover, for a general flat and homogeneous cosmological model
the luminosity distance can be expressed as an integral of the
Hubble function as follows:
\begin{eqnarray}\label{luminosity}
D_L (z) &=& {c\over H_0}(1+z)\int^{z}_{0}{1\over E(\zeta)}d\zeta,
\end{eqnarray}
where $E(z)= {H(z)\over H_0}$ is related to the Hubble
function expressed in terms of $z=a_0/a(t) - 1$. Let us note that
the luminosity distance depends also on the Hubble distance ${c
\over H_0}$ ( which does not depend on the choice of the unit of
time). Such \textit{freedom} allows us to fit $h$ or the \textit{a
priori} unknown age of the universe $\tau$ using the SNIa dataset.
We find that $\chi_{red}^2=0.98$ for 182 data points, and the best
fit values are ${\widehat H}_0=0.96^{+0.05}_{-0.05}$,
$s=-1.46^{+0.04}_{-0.03}$, which corresponds to $\Omega_{{\rm
\Lambda_{eff}}}=0.69^{+0.04}_{-0.06}$. We also get $h=0.72\pm 0.04$.

\begin{figure}
\centering{
 \includegraphics[width=5 cm, height=5
  cm]{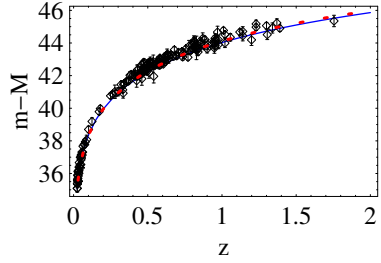}
        \caption{\small Observational data of the SNIa sample compiled by (\cite{Riess06})
        fitted to our model. The
    solid curve is the best fit curve, compared with a standard $\Lambda$CDM model
    with $\Omega_{\Lambda}=0.71$ (red dashed line). }
        \label{sup}}
\end{figure}

In Fig (\ref{sup}) we compare the best fit curve with the
observational data sets.

\subsubsection{Dimensionless
coordinate distance test} After having explored the Hubble diagram
of SNIa,  let us now follow the more general approach, suggested by
Daly \& Djorgovski (\cite{DD04}), and already tested in Paper I.
Consider as a cosmological observable the dimensionless coordinate
distance defined as \,:
\begin{equation}
y(z) = \int_{0}^{z}{\frac{1}{E(\zeta)} d\zeta} \,, \label{eq:
defy}
\end{equation}
noting that $y(z)$ does not depend explicitly on $h$, so that
any choice for $h$ does not alter the main result. Daly \&
Djorgovski have determined $y(z)$ for the SNIa in the Gold
Sample of Riess et al. (\cite{Riess04}) which is the most
homogeneous SNIa sample available today. Since SNIa allows to
estimate $D_L$ rather than $y$, a value of $h$ has to be set.
Fitting the Hubble law to a large set of low redshift ($z <
0.1$) SNIa, Daly \& Djorgovski (\cite{DD04}) have set:
\begin{displaymath}
h = 0. 66 \pm 0.08 .
\end{displaymath}
To enlarge the sample, Daly \& Djorgovski added 20  points on the
$y(z)$ diagram using a technique of distance determination based on
the angular dimension of radiogalaxies (\cite{DD04}). This data set
has been recently supplemented by 71 new supernovae from the
Supernova Legacy Survey of (\cite{SNLS}), which allowed the
determination of dimensionless coordinate distances to these
supernovae. These were obtained using the values and uncertainties
of $\mu_B$ listed in Table 9 of (\cite{SNLS}), with $h = 0.7$. This
extended sample that spans the redshift range $(0.1, 1.8)$ has been
suitably homogenized.

\begin{figure}
\centering{
        \includegraphics[width=6 cm, height=6 cm]{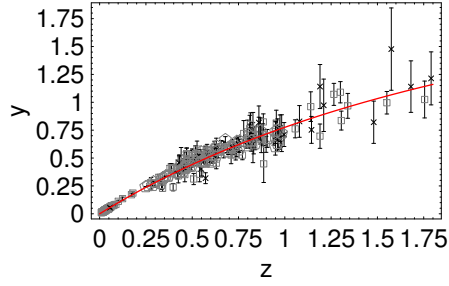}}
        \caption{\small Updated Daly \& Djorgovski database (\cite{DD05})
        fitted to our model. The solid curve is the best fit curve with
        $\chi_{red}^2=1.1$ for 248 data points,  and the best
        fit values are ${\widehat H}_0=.97^{+0.04}_{-0.03}$,
$s=-1.49^{+0.02}_{-0.04}$. }
        \label{cords}
\end{figure}

Using the following merit function\,:
\begin{equation}
\chi^2(s, {\widehat H}_0) = \frac{1}{N - 3} \sum_{i = 1}^{N}{\left
[ \frac{y(z_i;  \alpha_1, {\widehat H}_0) - y_i}{\sigma_i} \right
]^2}\,, \label{eq: defchi}
\end{equation}
we obtain that $\chi_{red}^2=1.1$ for 248 data points,  and the
best fit values are ${\widehat H}_0=.97^{+0.04}_{-0.03}$,
$s=-1.49^{+0.02}_{-0.04}$.  In Fig (\ref{cords})  we compare the
best fit curve with the observational data set, as in Paper I.
\subsubsection{H(z) and the relative galaxy ages}
\begin{figure}
\centering{
        \includegraphics[width=5cm, height=5 cm]{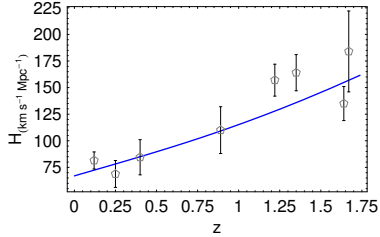}}
        \caption{\small The best fit curve of the measured values of $H(z)$  corresponds to
 ${\widehat H}_0=1.01^{+0.01}_{-0.03}$,
$s=-1.49^{+0.03}_{-0.09}$.}
        \label{jimenez-H2}
\end{figure}

In this section we discuss a possible observational determination of
$H(z)$ based on the method developed by (\cite{JVTS03}), that
involves differential age measurements. We present some constraints
that can be placed on the evolution of our quintessence model by
such  data. First, it is worth to stress some aspects connected with
the sensitivity of the cosmology to the $t(z)$ and ${dz \over dt}$
relations. Actually, it is well known that in scalar tensor theories
of gravity, as well as in general relativity, the expansion history
of the universe is determined by the function $H(z)$. This implies
that observational quantities, like the luminosity distance, the
angular diameter distance, and the lookback time, all depend on
$H(z)$.  It turns out that the most appropriate mathematical tool to
study the sensitivity of the cosmological model to such observables
is  the functional derivative of the corresponding relations with
respect to the cosmological parameters (see \cite{saini} for a
discussion about this point in relation to distance measurements).
\begin{figure}
\centering{ \includegraphics[width=5 cm, height=5
cm]{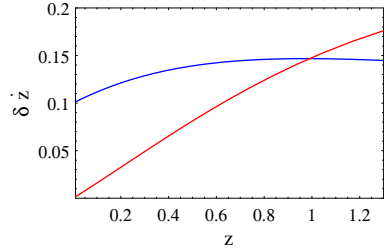}}\caption{\small We compare the sensitivity of the
${dz \over dt}$ relation to the values of the parameters in our
model. Actually  we plot the relative variation in ${dz \over dt}$
with respect to a variation of $s$ from $-1.4$ to $ -1.3$, and with
respect to a variation of $\widehat{H}_0$ from $1$ to\,$.9$, the
other parameters being fixed. The red line shows
$\displaystyle{\delta {dz \over dt}= {{dz \over
dt}(s=-1.3,H_0=1)-{dz \over dt}(s=-1.4,H_0=1)\over {dz \over
dt}(s=-1.4,H_0=1)}}$, and the blue line shows $\displaystyle{\delta
{dz \over dt}= {{dz \over dt}(s=-1.4,H_0=0.9)-{dz \over
dt}(s=-1.4,H_0=1)\over {dz \over dt}(s=-1.4,H_0=1)}}$.
}\label{comparez}
\end{figure}
\begin{figure}
\centering{ \includegraphics[width=5 cm, height=5
cm]{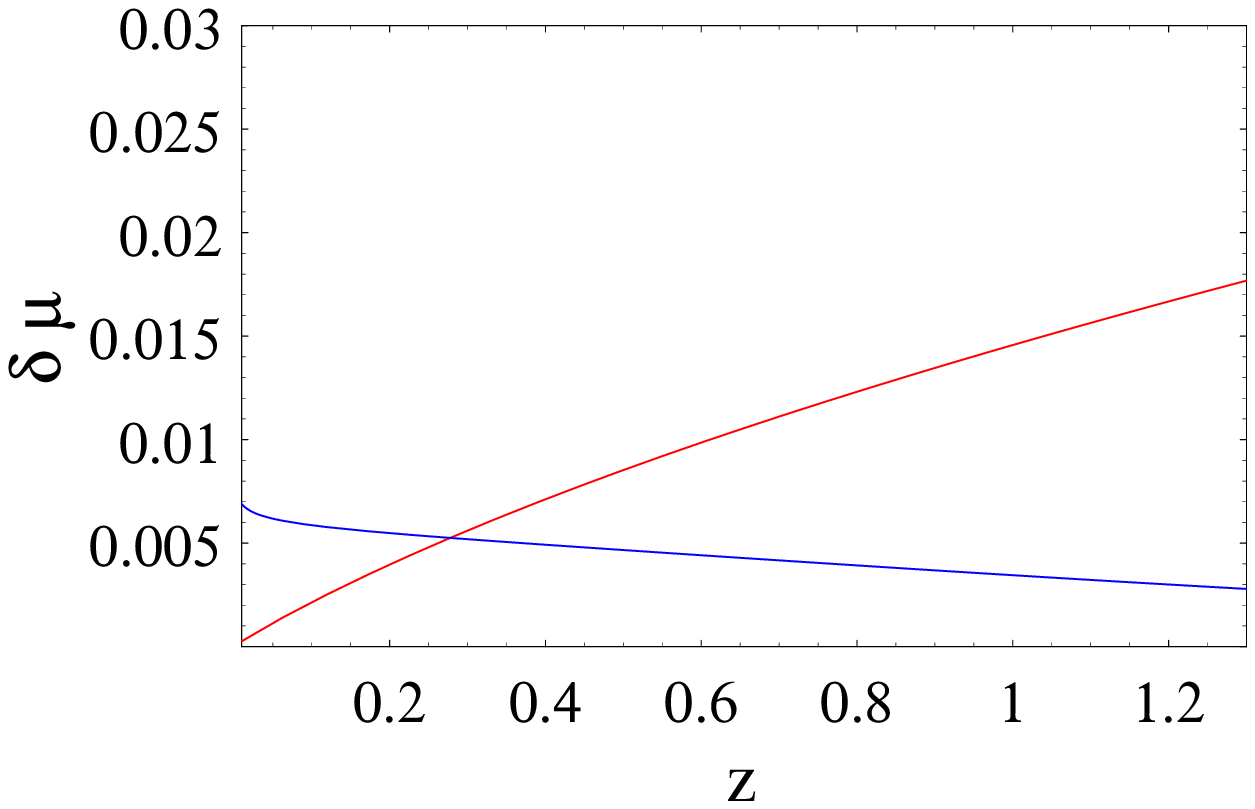}}\caption{\small We compare the sensitivity of the
\,$\mu=m-M$ relation to the values of the parameters in our model
in. Actually  we plot the relative variation in $\mu$ with respect
to a variation of $s$ from $-1.4$ to $ -1.3$, and with respect to a
variation of $\widehat{H}_0$ from $1$ to $.9$, the other parameters
being fixed. The red line shows $\displaystyle{\delta {d\mu \over
dt}= {{d\mu \over dt}(s=-1.3,H_0=1)-{d\mu \over
dt}(s=-1.4,H_0=1)\over {d\mu \over dt}(s=-1.4,H_0=1)}}$, and the
blue line shows $\displaystyle{\delta {d\mu \over dt}= {{d\mu \over
dt}(s=-1.4,H_0=0.9)-{d\mu \over dt}(s=-1.4,H_0=1)\over {d\mu \over
dt}(s=-1.4,H_0=1)}}$.  As wee see for the same variation in the
parameters the sensitivity of the distance modulus is quite smaller
than the respective variation in $ {dz\over dt}$ }\label{comparemu}
\end{figure}
However, also from an empirical point of view, it is possible to
show that the lookback time is much more sensitive to the
cosmological model than other observables, like the luminosity
distance, and the distance modulus. This circumstance encourages us
to use, together with other more standard techniques discussed
above, the age of \textit{cosmic clocks} to test alternative
cosmological scenarios.  Apart from the advantage of providing an
alternative instrument of investigation, the age-based methods use
the stronger sensitivity to the cosmological parameters of  the ${dz
\over dt}$ relation, as shown in Figs. (\ref{comparez}) and
(\ref{comparemu}). Moreover, as we will discuss in the following,
such a method reveals its full strength when applied to old objects
at very high $z$. Actually it turns out that this kind of analysis
could remove, or at least reduce, the degeneracy which we observe at
lower redshifts, for example the one in the Hubble diagram for SNIa
observations, which can be fitted by different cosmological models
with similar statistical significance.
%%\subsubsection{Differential ages of passively evolving galaxies}
Since the Hubble parameter can be related to the differential age of
the universe as a function of redshift by the equation
\begin{equation}
H(z)=- \frac{1}{1+z} \frac{dz}{dt}, \label{eq:hz}
\end{equation}
 a determination of $dz/dt$ directly measures $H(z)$.
Jimenez et al. ( \cite{JVTS03})  demonstrated the feasibility of
this method by applying it to a $z \sim 0$ sample of galaxies.
With the availability of new galaxy surveys it becomes possible to
determine $H(z)$ at $z > 0$.
 Here we use the  $dz/dt$ data from
(\cite{jim05}) to determine $H(z)$ in the redshift range $0.1 < z <
1.8$.  To follow the procedure described in (\cite{jim05}), first we
group together all galaxies that are within $\Delta z=0.03$ of each
other. This gives an estimate of the age of the universe at a given
redshift. We then compute age differences only for those bins in
redshift that are separated by more than $\Delta z= 0.1$ but less
than $\Delta z = 0.15$. The first limit is imposed so that the age
evolution between the two bins is larger than the error in the age
determination. We note here that differential ages are less
sensitive to systematic errors than absolute ages (see \cite{JVTS03}
). The observational value of $H(z)$ is then directly computed by
using Eq.~(\ref{eq:hz}), and after that the ages data have been
scaled according to our choice of the unit of time (the unknown
scaling factor has been provided by the $\chi^2$  procedure). To
determine the best fit parameters, we define the following merit
function\,:
\begin{equation}
\chi^2(s, {\widehat H}_0) = \frac{1}{N - 3} \sum_{i = 1}^{N}{\left
[ \frac{H(z_i;  s, {\widehat H}_0) - H_i}{\sigma_i} \right ]^2}\,.
\label{eq: defchih}
\end{equation}
We obtain that $\chi_{red}^2=1.09$ for 9 data points,  and the
best fit values are ${\widehat H}_0=1.01^{+0.01}_{-0.03}$,
$s=-1.49^{+0.03}_{-0.09}$. In Fig (\ref{jimenez-H2})  we compare
the best fit curve with the observational data set.
\begin{figure}
\centering{
        \includegraphics[width=5 cm, height=5 cm]{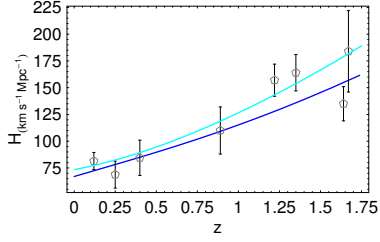}}
        \caption{\small The best fit curve to the $H(z)$
        data for our nmc model (dark blue line) and  for the quintessence QCDM
        fitted to the new released WMAP- three years + SNLS data
        (\cite{wmap})
        $\Omega_{\Lambda}=0.72\pm 0.04$,\, $\Omega_k=-0.010^{+0.0016}_{-0.0009}$,\,
        $w=-1.06^{+0.13}_{-0.08}$\, (blue line). It is interesting to note that future
        high redshift data could disentangle the degeneration among different models,
        since at high z the predicted $H(z)$ more sensitively
        depends on the values of the parameters.}
        \label{jimenezwmap}
\end{figure}

\subsection{The Sunyaev-Zeldovich/X-ray method}
In this section we discuss how the parameters of our model can be
constrained by the angular diameter distance $D_A$ as measured using
the Sunyaev-Zeldovich effect (SZE) and the thermal bremsstrahlung
(X-ray brightness data) for galaxy clusters. The distance
measurements using Sunyaev-Zeldovich effect and X-ray emission from
the intracluster medium have to take into account that these
processes depend on different combinations of some of the parameters
of the clusters (see \cite{birk} and references therein). The SZE is
a result of the inverse Compton scattering of the CMB photons on hot
electrons of the intercluster gas, which  preserves the number of
photons, but allows photons to gain energy and thus generates a
decrement of the temperature in the Rayleigh-Jeans part of the
black-body spectrum while an increment appears in the Wien region.
We limit our analysis to the so called {\it thermal} or {\it static}
SZE. The ${\it kinematic}$ effect, present only in clusters with a
nonzero peculiar velocity with respect to the Hubble flow along the
line of sight, will be neglected since typically the thermal SZE is
an order of magnitude larger than the kinematic one. As in Paper I,
we introduce the so called Compton parameter, $y$, defined as the
optical depth $\tau = \sigma_T \int n_e dl$ times the energy gain
per scattering:
\begin{equation}\label{compt}
  y=\int  \frac{k_B T_e}{m_e c^2} n_e \sigma_T dl,
\end{equation}
where $T_e$ is the temperature of the electrons in the
intracluster gas, $m_e$ is the electron mass, $n_e$ is the number
density of the electrons, and $\sigma_T$ is the Thompson cross
section of electron scattering, and the integration is performed along the line of
sight. In the low frequency regime of the
Rayleigh-Jeans approximation the shift of temperature is
\begin{equation}
 \frac{\Delta T_{RJ}}{T_0}\simeq -2y\,,
 \label{eq:sze5bis}
\end{equation}
where $T_0$ is the unperturbed CMB temperature. The next step to
quantify the SZE decrement is to specify the model for the
intracluster electron density and temperature distribution, which appropriately describes the observational properties of the gas.
Following (\cite{bonamente}) we use a hydrostatic equilibrium  double $\beta$-model model.
Actually, at the center of clusters the density may be high enough that the
radiative cooling time-scale is less than the cluster's age, leading
to a reduction in temperature and an increase in central density.
This can increase the central X-ray emissivity.  At large radii, the density of the gas is
sufficiently low that X-ray emission can be sustained for cosmological
periods without significant cooling. Therefore, cool core clusters effectively
exhibit two components: a centrally concentrated gas peak and a broad,
shallower distribution of the gas.  This phenomenon motivated the
modelling of the gas density with a function of the form:
\begin{equation}
n_e(r) = n_{e0} \cdot
\left[f\left( 1+\frac{r^2}{r_{c1}^2} \right)^{-\frac{3\beta}{2}}+
(1-f)\left( 1+\frac{r^2}{r_{c2}^2} \right)^{-\frac{3\beta}{2}} \right].
\label{density}
\end{equation}
The quantity $n_{e0}$ is the central density, $f$ governs the
fractional contributions of the narrow and broad components
($0\leq f \leq 1$), $r_{c1}$ and $r_{c2}$ are the two core radii
that describe the shape of the inner and outer portions of the
density distribution and $\beta$ determines the slope at large
radii (the same $\beta$ is used for both the central and outer
distribution in order to reduce the total number of degrees of
freedom) \footnote{It is worth to note that there are also
different hydrodynamical models which predict universal gas
density and gas temperature profiles that agree with the
observations (see for instance the one illustrated in
(\cite{seljak 2001}), derived from the universal dark matter
density profile, assuming that the gas density traces the dark
matter density in the outer parts of halos, or the one
introduced in (\cite{rasia})).}. This shape generalizes the
single $\beta$-model profile, introduced by Cavaliere and
Fusco-Femiano (1976) and commonly used to fit X-ray surface
brightness profiles, to a double $\beta$-model of the density
that has the freedom of following both the central spike in
density and the more gentle outer distribution. A double
$\beta$-model of the surface brightness was first used by Mohr
et al.\ (1999) to fit X-ray data of galaxy clusters; the density
model of Eq. (\ref{density}) was further developed by La~Roque
 (\cite{laroque2006}).
The X-ray surface brightness is related to the gas density as
\begin{equation}
S_X=\frac{1}{4 \pi (1+z)^4} \int n_e^2 \Lambda_{ee} dl
\label{XSB}
\end{equation}
where $z$ is the cluster redshift, $\Lambda_{ee}$ is the X-ray
cooling function, and it is a function of plasma temperature and
energy in the rest frame of the cluster, including contributions
from relativistic electron-ion thermal bremsstrahlung,
electron-electron thermal bremsstrahlung, recombination, and two
photon processes. The cluster angular diameter distance $D_A
\equiv dl/d\theta$, where $\theta$ is the line-of-sight angular
size, can be inferred with a joint analysis of SZE, taking
advantage of the different density dependence of the X-ray
emission and SZE decrement:
\begin{eqnarray}
\label{eq-system}
S_X \propto \int n_e^2 \Lambda_{ee} dl=\int n_e^2 \Lambda_{ee} D_A d\theta\\\nonumber
\Delta T_{CMB} \propto \int n_e T_e dl=\int n_e T_e D_A d\theta.
\end{eqnarray}
 Actually it turns out that
 \begin{equation}
D_A \propto \frac{\Delta T_{CMB}^2 \Lambda_{ee}}{S_X T_e^2}.
\label{eq-DA}
\end{equation}
Note that $D_A$ is
proportional to $\Delta T_{CMB}^2$ and $T_e^{3/2}$ (since
$\Lambda_{ee} \propto T_e^{1/2}$), so the
distance determination is strongly dependent on the accuracy of the
SZE decrement and X-ray temperature measurements.

\begin{figure}
\centering{
        \includegraphics[width=6 cm, height=5 cm]{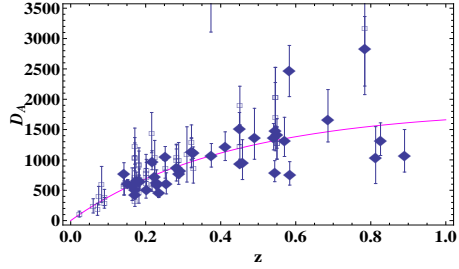}}
        \caption{\small Observational SZE data fitted to our model with
         the best fit values  ${\widehat
H}_0=1^{+0.01}_{-0.03}$, $s=-1.49^{+0.03}_{-0.09}$, and $h=0.70\pm
0.05$. The empty boxes indicate distance
measurements for a sample of 44 mentioned clusters
 (see \cite{birk, reese}),while the filled diamonds indicate the measurement
 of the angular diameter distances  from {\it Chandra} X-ray imaging
and Sunyaev-Zel'dovich effect mapping of 39 high redshift clusters
(\cite{bonamente}).}
        \label{szall}
\end{figure}

Recently distances to 18 clusters with redshift ranging from $z\sim
0.14$ to $z\sim 0.78$ have been determined from a likelihood joint
analysis of SZE and X-ray observations (see Table 7 in
\cite{reese}). We perform our analysis using angular diameter
distance measurements for a sample of 83 clusters, containing the 18
above mentioned clusters,  other 24 known previously (see
\cite{birk}), and a recently released sample with the measurement of
the angular diameter distances  from the {\it Chandra} X-ray imaging
and Sunyaev-Zel'dovich effect mapping of 39 high redshift clusters
of galaxies ($0.14 \leq z \leq 0.89$ )\,(\cite{bonamente}). The
unprecedented spatial resolution of \textit{Chandra} combined with
its simultaneous spectral resolution allows a more accurate
determination of  distances. Let us consider the merit function of
the form\,:
\begin{eqnarray}
\label{eq: defchitot}
 &&\chi^2(s, {\widehat H}_0) = \frac{1}{N - 3} \sum_{i =
1}^{N}\left[ \frac{(D_A(z_i; s, {\widehat H}_0)-D_i) }{\sigma_i}
\right]^2.
\end{eqnarray}
Fitting the data we obtain that $\chi_{red}^2=1.2$ for 83
data points, and the best fit values are ${\widehat
H}_0=1^{+0.01}_{-0.03}$, $s=-1.49^{+0.03}_{-0.09}$ and $h=0.70\pm
0.05$. In Fig. (\ref{szall}) we compare the best fit curve with the
observational SZE data.
\subsection{Gamma-Ray Burst Hubble Diagram }
\begin{figure}
\centering{
 \includegraphics[width=5 cm, height=5 cm]{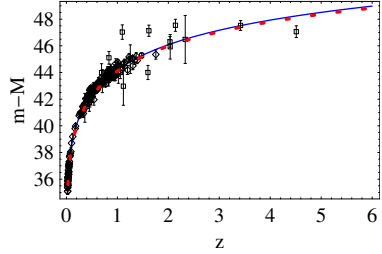}}
        \caption{\small Observational Hubble diagram for the recent SNIa sample
        compiled by (Riess et al. 2006) (empty lozenges), and the GRBs
         data (\cite{ BFK2003}) (empty boxes) fitted to our model. The
         solid curve is the best fit curve with,\,${\widehat H}_0=\, 0.98^{+0.03}_{-0.03}$,\,
$s=-1.43^{+0.02}_{-0.04}$. The
red dashed line corresponds to the standard $\Lambda$CDM model
with $\Omega_{\Lambda}=0.71$}
        \label{hubblediagram}
\end{figure}

Gamma-ray bursts (GRBs) are bright explosions visible across
most of the Universe, certainly out to redshifts of $z \sim 7$
and likely out to $z \sim 10$. Recent studies have pointed out
that GRBs may be used as standard cosmological candles
(\cite{ghirlanda, FB05}): actually it turns out that the energy
released during bursts spans nearly three orders of magnitude,
and the distribution of the opening angles of the emission, as
deduced from the timing of the achromatic steepening of the
afterglow emission, spans a similar wide range of values.
However, when the apparently isotropic energy release and the
conic opening of the emission are combined to infer the
intrinsic, true energy release, the resulting distribution does
not widen, as is expected for uncorrelated data, but shrinks to
a very well determined value (\cite{fk03}), with a remarkably
small (one--sided) scattering, corresponding to about a factor
of $2$ in total energy. Similar studies in the X--ray band have
reproduced the same results. It is thus very tempting to study
to what extent this property of GRBs makes them suitable
cosmological standard candles. Schaefer (\cite{schaefer})
proposed to use the two well known correlations of the GRBs
luminosity (with variability, and with time delay), while others
exploited the recently reported relationship between the
beaming--corrected $\gamma$-ray energy and the locally observed
peak energy of GRBs (see for instance \cite{day}). As for the
possible variation of ambient density from burst to burst, which
may widen the distribution of bursts energies, Frail \& Kulkarni
(\cite{fk03}) remarked that this spread is already contained in
their data sample, and yet the distribution of energy released
is still very narrow. There are at least two reasons  why GRBs
are better than type Ia supernovae  as cosmological candles. On
the one hand, GRBs are easy to find and locate: even 1980s
technology allowed BATSE to locate $\sim$1 GRB per day, making
the build--up of a 300--object database a one--year enterprise.
The {\it Swift} satellite launched on 20 November 2004,  detects
GRBs at about the same rate as BATSE, but with a nearly perfect
capacity for identifying their redshifts simultaneously with the
afterglow observations \footnote{http://swift.gsfc.nasa.gov}.
Second, GRBs have been detected out to very high redshifts: even
the current sample contains several events with $z> 3$, with one
(GRB 000131) at $z = 4.5$ and another at $z=6.3$. This should be
contrasted with the difficulty of locating SN at $z > 1$, and
absence of any SN with $z
> 2$. On the other hand, the distribution of luminosities of SNIa is
narrower than the distribution of energy released by GRBs,
corresponding to a magnitude dispersion $\sigma_{M(SN)} = 0.18$
rather than $\sigma_{M(GRB)} = 0.75$. Thus GRBs may provide a
complementary standard candle, out to distances which cannot be
probed by SNIa, their major limitation being the larger intrinsic
scatter of the energy released, as compared to the small scatter in
peak luminosities of SNIa. There currently exists enough information
to calibrate luminosity distances and independent redshifts for nine
bursts (\cite{schaefer}). These bursts were all detected by BATSE
with redshifts measured from optical spectra of either the afterglow
or the host galaxy.  The highly unusual GRB980425 (associated with
supernova SN1998bw) is not included because it is likely to be
qualitatively different from the classical GRBs. Bursts with red
shifts that were not recorded by BATSE cannot yet have their
observed parameters converted to energies and fluxes that are
comparable with BATSE data. For the present analysis we shall use a
sample of GRBs that had their redshifts estimated (\cite{ BFK2003}),
as represented by empty boxes in Fig. (\ref{hubblediagram}), with
the distance modulus $\mu$, given by Eq. (\ref{eq:modg}).

To this aim, the only difference with respect to the SNIa is that
we {\it slightly} modify the {\it correction term} of Eq.
(\ref{corgef}), and we take
\begin{equation}\label{corgef2}
\Delta m_{G_{eff}}=2.5\gamma \frac{\Delta G_{eff}(t)}{\left( \ln
10\right) G_{eff}}.
\end{equation}
We expect that $\gamma $ is of order unity, so that the
$G$-correction would be roughly half a magnitude. We obtain
$\chi_{red}^2=1.1$ for $24$ data points, and the best fit value
is\,${\widehat H}_0=\, 0.98^{+0.03}_{-0.03}$,\,
$s=-1.43^{+0.02}_{-0.04}$, which are compatible with the SNIa
results. We also find that $\gamma$ which appears in Eq.
(\ref{corgef2}) is equal to $1.5$. In Fig. (\ref{hubblediagram})
we compare the best fit curve with both the GRBs and the SNIa
Gold Sample.
\subsection{The gas fraction in clusters}
\begin{figure}
\centering{
        \includegraphics[width=5cm, height=6 cm]{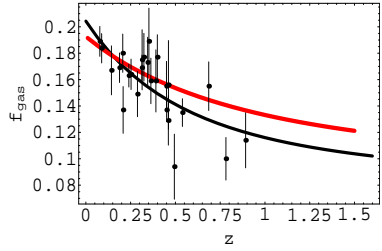}}
        \caption{\small In the diagram we plot the best fit curve to the $f_{gas}$
        data for our nmc model (red thick line) and  for the quintessence model (black thick line)
        considered in (\cite{limagas}).  It is interesting to note that, as pointed out also for the model described in (\cite{nmc}),
      even if the statistical significance of the best
        fit procedure for these two models is comparable, the best fit relative to our nmc model seems
        to be dominated by smaller redshift data,
        while the one relative to the Lima \& al. model by higher redshift data.  }
        \label{fgascomp}
\end{figure}

Measurements of the gas mass fraction in galaxy clusters have been
proposed as a test of cosmological models (\cite{fgasdata}). Both
theoretical arguments and numerical simulations predict that the
baryonic mass fraction in the largest relaxed galaxy clusters should
not depend on the redshift, and should provide an estimate of the
cosmological baryonic density parameter $\Omega_b$ (\cite{ENF98}).
The baryonic content in galaxy clusters is dominated by the hot
X\,-\,ray emitting intra-cluster gas so that what is actually
measured is the gas mass fraction $f_{gas}$ and it is this quantity
that should not depend on the redshift. Moreover, it is expected
that the baryonic mass fraction in clusters is equal to the
universal ratio $\Omega_b/\Omega_m$ so that $f_{gas}$ should indeed
be given by $b {\cdot} (\Omega_b/\Omega_m)$, where  the
multiplicative factor $b$ is motivated by simulations that suggest
that the gas fraction is lower than the universal ratio. Following
the procedure described in (\cite{fgasdata,allen2}), and already
used in Paper I we adopt the standard CDM model (i.e., a flat
universe with $\Omega_m = 1$ and $h = 0.5$) as a reference cosmology
in making the measurements so that the theoretical expectation for
the apparent variation of $f_{gas}$ with the redshift is:
\begin{equation}
f_{gas}(z) = \frac{b \Omega_b}{(1 + 0.19 \sqrt{h}) \Omega_m} \left
[ \frac{D_A^{SCDM}(z)}{D_A^{mod}(z)} \right ]^{1.5}\,, \label{eq:
fgas}
\end{equation}
where we substitute the appropriate expression of $\Omega_m$ for
our model, and $D_A^{SCDM}$ and $D_A^{mod}$ are the angular
diameter distance for the SCDM and our model respectively. Allen
\& al. (\cite{fgasdata}) have extensively analyzed the set of
simulations in\,(\cite{ENF98}) to get $b = 0.824 {\pm} 0.089$,
so in our analysis below, we  set $b = 0.824$. Actually, we have
checked that, for values in the $2 \sigma$ range quoted above,
the main results do not depend on $b$. Moreover we have defined
the following merit function\,:
\begin{equation} \chi^2 =
\chi_{gas}^2 + \left(\frac{\Omega_{\rm b}h^2-0.0214}{0.0020}
\right)^2 +\left(\frac{h-0.72} {0.08}
\right)^2+\left(\frac{b-0.824} {0.089} \right)^2\,, \label{eq:
defchin}
\end{equation}
where
\begin{equation}
\chi_{gas}^2 = \sum_{i = 1}^{N_{gas}}{\left [ \frac{f_{gas}(z_i,
\alpha_1,{\widehat H}_0) - f_{gas}^{obs}(z_i)}{\sigma_{gi}} \right
]^2}. \label{eq: chigas}
\end{equation}
Here $f_{gas}^{obs}(z_i)$ is the measured gas fraction in galaxy
clusters at redshift $z_i$ with an error $\sigma_{gi}$ and the
sum is over the $N_{gas}$ clusters considered. Let us note that
recently Allen \& al. (\cite{allen2}) have released a catalog of
26 large relaxed clusters with a precise measurement of both the
gas mass fraction $f_{gas}$ and the redshift $z$. We use these
data to perform our likelihood analysis, we get $\chi^2=1.16$
for 26 data points, and $s=-1.39^{+0.04}_{-0.01}$, ${\widehat
H}_0=0.98\pm 0.04$, $h=0.65\pm 0.05$, and $w_\phi=-0.95\pm
0.07$. A brief comparison of our results with similar recent
results of Lima et al. (\cite{limagas}), where the equation of
state characterizing the dark energy component is constrained by
using galaxy clusters x-ray data, can still be done. In their
analysis, however, they consider quintessence models \textit{ in
the standard gravity theories}, with a non evolving equation of
state, but they allow the so-called phantom dark energy with $w
< -1$, which violates the null energy condition. As the best
fit value of $w$ to the data of (\cite{fgasdata}) they obtain
$w=-1.29_{-0.792}^{+0.686}$. In order to directly compare this
result with our analysis we first fit the model considered in
(\cite{limagas}) to the updated and wider data set of
(\cite{allen2}), used in our analysis. To this aim we also refer
to the model function $f_{gas}(z)$, and the merit function
$\chi^2$, defined in the Eqs. (\ref{eq: fgas}) and (\ref{eq:
defchin}) respectively. We get $\chi^2=1.175$ for 26 data
points, and $\Omega_m=0.23^{+0.05}_{-0.03}$,
$h=0.76^{+0.04}_{-0.09}$, and $w=-1.11\pm 0.35$, so $w<-1$, what
corresponds to a phantom energy. Let us note that our model,
instead, gives $w_\phi=-0.87\pm 0.05$, what does not violate the
null energy condition.\, In Fig. (\ref{fgascomp})  we compare
the best fit curves for our and the Lima \& al. model with the
observational data.

\section{Growth of density perturbations}
\begin{figure}
\centering{
        \includegraphics[width=5 cm, height=6 cm]{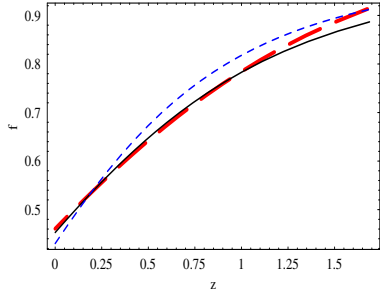}}
        \caption{\small The growth index $f$ in different cosmological
        models.
        The thick dashed red line corresponds to  our non minimally
        coupled model. The blue thin dashed curve corresponds to the standard  $\Lambda$-CDM model with $\Omega_m=0.25$, and the
        black solid line corresponds to another quintessence
        model with an exponential potential (described in \cite{pv}).}
        \label{peculiarcompare}
\end{figure}

In this section we consider the evolution of scalar density
perturbations in the longitudinal gauge $ds^2= -(1 + 2 \Phi) dt^2 +
a^2 (1 - 2\Phi) d{\bf x}^2$.  Contrary to what happens in the
framework of the minimally coupled theory, where we have to deal
with a fully relativistic component, which becomes homogeneous on
scales smaller than the horizon -- so that standard quintessence
cannot cluster on such scales -- in the non minimally coupled
quintessence theories it is possible to separate a pure
gravitational term both in the stress-energy tensor $T_{\mu \nu}$,
and in the energy density $\rho_\phi$, so the situation changes, and
it is necessary to consider also fluctuations of the scalar field.
However, it turns out (\cite{beps, uzan2}) that the equation for
dust like matter density perturbations inside the horizon can be
written as follows:
\begin{equation}
{\ddot \delta_m} + 2H {\dot \delta_m} - {1\over 2} G_{\rm Cav}\,
\rho_m~\delta_m\simeq 0~,\label{del}
\end{equation}
where $G_{\rm Cav}$ is the effective gravitational constant
defined by
\begin{equation}
G_{\rm Cav}=
\,-\frac{1}{2\xi(s)\phi^2}\left[\frac{1+16\xi(s)}{1+12\xi(s)}\right]\,.\label{Geff}
\end{equation}
The equation (\ref{del}) describes, in the non minimally coupled
models, evolution of the CDM density contrast, $\delta_m \equiv
\delta \rho_m /\rho_m$, for perturbations inside the horizon. In
our model the Eq.(\ref{del}) is rather complicated and takes the
form
\begin{eqnarray}\label{pertdef}
&& {\ddot \delta_m} + 2
\left[\frac{t^{\frac{1}{s+3}-2-\frac{1}{s}} \left((s (2 s+9)+6)
B(s) t^{\frac{3}{s+3}}+(2 s
   (s+3)+3) \left(\left(\frac{1}{A(s)}\right)^{\frac{s}{s+1}}-B(s)\right)\right)
   \left(\left(\frac{1}{A(s)}\right)^{\frac{s}{s+1}}+\left(t^{\frac{3}{s+3}}-1\right)
   B(s)\right)^{-1-\frac{1}{s}}}{A(s)}\right]\,{\dot \delta_m}\,\nonumber -\\
   &&\frac{48 (s+1) (s+2)
   t^{\frac{\left(s+\frac{3}{2}\right)^2}{4 s (s+3)}}
   \left(\left(\frac{1}{A(s)}\right)^{\frac{s}{s+1}}+B(s)
   \left(-\frac{12 (s+1) (s+2)V
   t^{\frac{3}{s+3}}}{2
   s+3}-1\right)\right)^{2+\frac{3}{s}}}{2 C(s)^2 (2
   s+3)^2} \left({1+16 \xi(s)\over 1+12 \xi(s)}\right)\rho_m ~\delta_m\simeq
   0\end{eqnarray}
Eq. (\ref{pertdef}) does not admit exact solutions, and can be
solved only numerically. However, since with our choice of
normalization the whole history of the Universe is confined to
the range $t\in[0,1]$ and therefore to study the behavior of
solutions for $t\simeq 0$ we can always expand the functions in
Eq.~(\ref{pertdef}) in series around $t=0$, in order to get
approximate solutions. Actually we obtain an integrable Fuchsian
differential equation, which is a hypergeometric equation. We
then use such a solution to set the initial conditions at $t=0$
to numerically integrate Eq. (\ref{pertdef}) in the whole range
$[0,1]$. We use the growing mode $\delta_+$ and define the
growth index $f$ as
\begin{equation}
\lab{grow} f \equiv \frac{d \ln \delta_+ }{d \ln a}\,,
\end{equation}
where $a$ is the scale factor.
 Once we know how the growth index $f$ evolves with  redshift
and how it depends on our model parameters, we can use the
available observational data to estimate the values of such
parameters, and the present value of $\Omega_{{\rm m}}$. The
2dFGRS team has recently collected positions and redshifts of
about $220 000$ galaxies and presented a detailed analysis of the
two-point correlation function. They measured the redshift
distortion parameter $\beta=\displaystyle{f\over b}$, where $b$ is
the bias parameter describing the difference in the distribution
of galaxies and mass, and obtained that $\beta_{|z\rightarrow
0.15}=0.49 \pm 0.09$ and $b=1.04 \pm 0.11$. From the
observationally determined $\beta$ and $b$ it is now
straightforward to get the value of the growth index at $z=0.15$
corresponding to the effective depth of the survey. Verde \&
al.~(\cite{ver+al01}) used the bispectrum of 2dFGRS galaxies, and
Lahav \& al. ~(\cite{la+al02}) combined the 2dFGRS data with CMB
data, and they obtained
\begin{eqnarray}\label{bias}
 b_{verde}&=&1.04\pm 0.11\,,\\
 b_{lahav}&=&1.19\pm 0.09\,.
\end{eqnarray}
Using these two values for $b$ we calculated the value of the
growth index $f$ at $z=0.15$, we get respectively
\begin{eqnarray}\label{peculiar}
 f_1&=&0.51\pm 0.1\,,\\
 f_2&=&0.58 \pm 0.11\,.
\end{eqnarray}
To evaluate the growth index at $z=0.15$ we first have to invert the
$z-t$ relation and find $t(0.15)$: actually the $z-t$ relation is
rather involved and cannot be analytically inverted, so we perform
this inversion numerically. Finally, we get $s=-1.4\pm 0.1$$,
{\widehat H}_0=0.94^{+0.05}_{-0.02}$,\,\,\,$V_0=0.5\pm 0.06$ which
corresponds to $\Omega_{\Lambda_{eff}}=0.65\pm 0.08$. In Fig.
(\ref{peculiarcompare}) we show how the growth index is changing
with redshift in our non minimally coupled model as compared with
the standard $\Lambda$-CDM model, with $\Omega_m=0.25$, and a
quintessence model namely the minimally coupled exponential model
described in (\cite{pv}). We note that at low redshift theoretical
predictions of these different models are not distinguishable,
independent measurements from large redshift surveys at different
depths could disentangle this degeneracy.
\section{Summary and conclusions}

In this paper we have extended the analysis that we performed in
Paper I (where  we have analyzed a special extended quintessence model, based on one of the most
commonly used quintessence potentials $V(\phi)=\lambda \phi^{4}$,
corresponding to the coupling $F(\phi)=(3/32)\phi^2$ ), considering a new and wider class of
theories for which exact solutions of the Einstein
equations are known, and discussed how in such models it is possible
to treat the fine tuning problem in an alternative way. We have shown that in the family of such models
selected by requiring that their corresponding point like Lagrangian
admits a Noether symmetry an epoch of accelerated expansion appears
in a natural way. In the non minimally coupled scalar tensor theory
of gravity it is possible to perform an appropriate conformal
transformation and to move from the Jordan picture to the standard
Einstein one but then matter becomes coupled to the scalar field. We
have explored both descriptions and also considered the neutrino
mass varying model as a possible example of non minimally coupled
scalar tensor theory.

\begin{figure}
\begin{minipage}{13 cm}
\centering{
       \includegraphics[width=6 cm, height=5 cm]{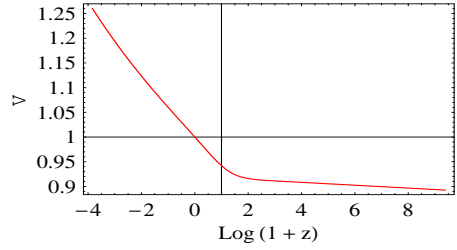}}
      \caption{\small The potential $V$ as a function of the redshift $z$.}
       \label{Vvariation1}
\end{minipage}
\end{figure}

It turns out that the imposed requirement of existence of a Noether
symmetry is quite restrictive and we obtained a family of models
that is fully specified by 3 parameters: a parameter $s$ that
determines the strength of the non minimal coupling and the
potential of the scalar field, $H_0$ the Hubble constant, and a
parameter $V_0$ that determines the scale of the potential. To
determine the values of these parameters we compared predictions of
our model with several independent observational data. The results
of this parameter determination procedure are presented in Table 2.
We see that with our average value of $s= - 1.46$ the scale factor,
for small $t$, is changing as $a\sim t^{0.66}\sim t^{2/3}$, and for
large $t$, as $a\sim t^{1.25}\sim t^{5/4}$ while $\phi\sim
t^{0.0032}$ and $\phi \sim t^{0.083}$ in corresponding asymptotic
regimes. The potential $V$ is decaying to zero for large $t$, after
reaching a maximum value, (see Fig. (\ref{Vvariation1})). Similarly,
the effective gravitational coupling $G_{eff}=-{1\over {2F}}$ is
decreasing for  large $t$, until it becomes zero for $t\rightarrow \infty
$ ( we have a sort of \textit{ asymptotic freedom } at $
t\rightarrow \infty $). It turns out that in our model the
observational constraints on the variation of the effective
gravitational constant are respected. Actually a new analysis of the
Big Bang Nucleosynthesis~(\cite{bbnG}) restricts the variations of
$G$ to
\begin{equation}
 \Delta G_{BBN}/G_0 = 0.01^{+0.20}_{-0.16}
\end{equation}
at $68\%$ confidence level, where  $\Delta G_{BBN}/G_0
={G_{BBN}-G_{0}\over G_0} $, and $G_0\equiv G_{Newton}$. The
combined analysis of the new $^4$He and WMAP data implies that
\begin{equation}
 -0.10 < \Delta G_{BBN}/G_0 < 0.13\,.
\end{equation}
A recent analysis of the secular variation of the period of
nonradial pulsations of the white dwarf G117-B15A
shows~(\cite{gwd}) that $0<\dot G/G < 4.0\times 10^{-11}\,{\rm
yr}^{-1}$ at 2$\sigma$, which is of the same order of magnitude as
previous independent bounds (see also \cite{gwd2}). With our unit
of time this becomes $0<\,\dot G/G < 4.0 \,\tau \times 10^{-2}$,
where  $\tau = 13.73^{+ 0.16}_{-0.15}$  as given by the WMAP team (\cite{spergel}).

Initially, for small $t$, the matter energy density is larger than the energy
density of the scalar field.

In Table (\ref{tab1}) we present results of our analysis, they
show that predictions of our model are fully compatible with the
recent observational data. Comparing results of this paper with
our previous analysis of minimally coupled scalar field models
(see Paper I) we conclude that the present day observational
data connected with the post recombination evolution of the
universe can be fitted by several different models of
quintessence. More data on high redshift supernovae of type Ia
and GRBs are needed as well as more information on the early
phase of structure formation in order to place stronger
restrictions on the allowed type of dark energy.
\begin{table*}[ht]\label{tab1}
\begin{center}
\caption{The basic cosmological parameters derived from our model
are compared with observational data. }
\begin{tabular*}{12.15 cm}{|c|c|c|c|c|}
  \cline{1-5}
  \bf{dataset} & $\mathbf{s}$ & $\mathbf{{\widehat H}_0}$  & $\mathbf{\Omega_{\Lambda_{eff}}}$ & $w_{\phi}$ \\
  \hline
  high redshift SNIa  &  $ -1.46^{+0.04}_{-0.03}$ & $ 0.96^{+0.05}_{-0.05}$& $0.69^{+0.04}_{-0.06}$ & $-1.01\pm 0.02$  \\
\hline
  dimensionless coordinates test &  $-1.49^{+ 0.02}_{-0.04}$ & $ 0.97^{+0.04}_{-0.03}$ & $ 0.73\pm 0.08 $ &$-0.98 \pm 0.03$ \\
\hline
  relative galaxy ages &  $-1.49^{+0.03}_{-0.09}$ & $ 1.01^{+0.01}_{-0.03}$ & $ 0.80 \pm 0.07$ & $-0.99 \pm 0.04$\\
\hline   sze data &     $-1.49^{+0.03}_{-0.05}$&$ 1^{+0.01}_{-0.03}$ &$ 0.70\pm  0.05$&$ -1 \pm 0.04$  \\
\hline  GRBs dataset &$-1.43^{+0.02}_{-0.04}$ & $0.98^{+0.03}_{-0.03}$ & $ 0.76 \pm 0.06 $  &$-0.9\pm 0.04 $  \\
\hline  fraction of gas in clusters & $-1.39^{+0.04}_{-0.01}$&$ 0.98\pm 0.04$ &$ 0.77^{+0.05}_{-0.03}$&$-0.87 \pm 0.05$  \\
\hline galaxies peculiar velocity &   $-1.4\pm
0.1$&$0.94^{+0.05}_{-0.02}$ &$ 0.65\pm 0.08 $&$-0.85 \pm 0.04$  \\
\hline average &   $\mathbf{-1.46 \pm 0.09}$&$\mathbf{ 0.97 \pm
0.02}$ &$ \mathbf{0.74\pm 0.03}$&$\mathbf{-1.01\pm 0.02}$ \\
\hline
\end{tabular*}
\end{center}
\end{table*}

\section*{Acknowledgments}
This work was supported in part by the grant of Polish Ministry of
Science and Higher Education 1-P03D-014-26, and by INFN Na12. The
authors are very grateful to Professor Djorgovski, for providing the
data that we used in Sec. 3.1.1, and to Professors Verde and Simon,
for providing the data used in Sec. 3.2.2. Of course we take full
responsibility of the fitting procedure.

\end{document}